\documentclass[nofootinbib,floats,aps,groupedaddress,preprint]{revtex4}
\usepackage{amssymb,amsmath,amsbsy}
\usepackage{mathrsfs,verbatim,enumerate}
\usepackage[dvips]{graphics}
\usepackage{epsfig}

\def\beq{\begin{equation}}
\def\eeq{\end{equation}}
\def\beqa{\begin{eqnarray}}
\def\eeqa{\end{eqnarray}}
\def\ifmath#1{\relax\ifmmode #1\else $#1$\fi}
\def\lsup#1{^{\lower 4pt\hbox{$\scriptstyle#1$}}}
\def\llsup#1{^{\lower 2pt\hbox{$\scriptstyle#1$}}}

\def\lsim{\mathrel{\raise.3ex\hbox{$<$\kern-.75em\lower1ex\hbox{$\sim$}}}}
\def\gsim{\mathrel{\raise.3ex\hbox{$>$\kern-.75em\lower1ex\hbox{$\sim$}}}}

\def\fig#1{Fig.~\ref{#1}}
\def\eq#1{eq.~(\ref{#1})}

\begin{document}
\pagestyle{empty}
\begin{center}
{\Large\bf General Analysis of Antideuteron Searches for Dark Matter}

\vspace{1cm}

{\sc Yanou Cui,$^{a,}$}\footnote{E-mail:  ycui@physics.harvard.edu}
{\sc John D. Mason,$^{a,}$}\footnote{E-mail: jdmason@physics.harvard.edu}
{\sc\small and}
{\sc Lisa Randall$^{a,}$}\footnote{E-mail: randall@physics.harvard.edu}
\vspace{0.5cm}

{\it\small $^a$Jefferson Physical Laboratory, Harvard University,
Cambridge, Massachusetts 02138, USA}

\end{center}

\vspace{0.5cm}
\begin{center}
\today
\vspace{1cm}

\begin{abstract}
Low energy cosmic ray antideuterons provide a unique low background channel for indirect detection of dark matter. We compute the cosmic ray flux of antideuterons from hadronic annihilations of dark matter for various Standard Model final states and determine the mass reach of two future experiments (AMS-02 and GAPS) designed to greatly increase the sensitivity of antideuteron detection over current bounds. We consider generic models of scalar, fermion, and massive vector bosons as thermal dark matter, describe their basic features relevant to direct and indirect detection, and discuss the implications  of direct detection bounds on models of dark matter as a thermal relic. We also consider specific dark matter candidates and assess their potential for detection via antideuterons from their hadronic annihilation channels. Since the dark matter mass reach of the GAPS experiment can be well above $100~{\rm GeV}$, we find that antideuterons can be a good indirect detection channel for a variety of thermal relic electroweak scale dark matter candidates, even when the rate for direct detection is highly suppressed.
\end{abstract}

\end{center}
 \maketitle

\newpage
\setcounter{page}{1}
\setcounter{footnote}{0}
\pagestyle{plain}

\tableofcontents
\vfill\eject

\section{Introduction To Antideuterons}

Antimatter is rare in our Universe, making it potentially a good low-background channel in which to search for evidence that dark matter annihilates to Standard Model particles. Due to color multiplicity, it is reasonable to expect a sizable fraction of annihilations into hadronic final states--either directly or through subsequent decay.  Dark matter annihilation to hadronic final states can account for the antiproton ($\bar{p}$) cosmic ray spectrum, which also provides constraints. This component has been calculated for various models of dark
matter \cite{Bottino:1998tw,Bergstrom:1999qv,Donato:2003xg,Barrau:2005au,Grajek:2008jb,Cirelli:2008id,Cirelli:2009uv,Nezri:2009jd}. And antiproton
cosmic ray data  \cite{Orito:1999re,Maeno:2000qx,Haino:2004nq,Fuke:2005it,Adriani:2008zq} constrains the dark matter's mass and annihilation rate for models with antiprotons in the final state. For instance, for dark matter with a mass $m_{DM}< 50~{\rm GeV}$, $\langle \sigma |v| \rangle_{therm} \sim 1 ~{\rm pb}$ (cross section required by thermal relic density), and annihilation to $b\bar{b}$, one finds that the dark matter contribution to the $\bar{p}$ spectrum can exceed experimental bounds \cite{Bottino:1998tw,Donato:2003xg,Donato:2008jk}. On the other hand, for dark matter masses of $m_{DM}  \geq 50~{\rm GeV}$ and an annihilation cross section $\langle \sigma |v| \rangle_{ann} \sim 1~{\rm pb} $, the primary contribution to the antiproton cosmic ray flux of dark matter annihilation will generally be smaller than the secondary contribution from astrophysical processes. Therefore, only if the antiproton cosmic ray flux is known to a high degree of precision would it be possible to disentangle a dark matter signal from the astrophysical background.

Compared to antiprotons, antideuteron  cosmic rays are poised to be a more sensitive indirect probe of dark matter with
a hadronic annihilation channel \cite{Donato:1999gy,Baer:2005tw,Donato:2008yx,Ibarra:2009tn,Braeuninger:2009pe}. The threshold energy required to create an
antideuteron from cosmic ray proton collisions with interstellar H and He is $E_{th} = 17 {\rm m_p}$ \footnote{A $pp$ collision which makes an antideuteron must have $\hat{s} > 36m_p^2$. In the rest frame of a Hydrogen nucleus in the galactic disk $\hat{s} = (m_p + E_p)^2 -k_p^2$, where $E_p$ and $k_p$ are the energy and momentum of an impinging cosmic ray proton. This implies that $E_p > 17m_p$ in order for this process to occur.} (for
antiprotons it is $E_{th}= 7 {\rm m_p}$). Furthermore, the low value of the antideuteron binding energy ($B =
2.2 {\rm MeV}$), removes its ability to slow down as it propagates in the galactic disk, leading to a low
background for low energy ($T_{\bar{D}/n} < 1 {\rm GeV}$) antideuterons having an astrophysical origin
\cite{Chardonnet:1997dv,Donato:1999gy,Duperray:2005si}. The fact that antideuterons can act as a more sensitive probe of the hadronic final states of dark matter annihilation was first investigated in \cite{Donato:1999gy} where an antideuteron signal originating from supersymmetric dark matter was investigated.

Future experiments, two energy bands of the Alpha Magnetic Spectrometer Experiment (AMS-02) \cite{Ahlen:1994ct,ams02} and the three \footnote{ The three proposals are the Long Duration  Balloon(LDB), the Ultra-Long Duration Balloon(ULDB), and a satellite (SAT) mission.}  phases of the General Antiparticle Spectrometer (GAPS) \cite{Mori:2001dv, Fuke:2008zz}, should greatly increase the sensitivity of searches for cosmic ray antideuterons over the current bound on the antideuteron component of the cosmic ray Spectrum set by BESS  \cite{Fuke:2005it}. The current bound is:
\beqa
&&\Phi_{\bar{D}}  < 0.95 \times 10^{-4} [m^2~s~sr~{\rm GeV}]^{-1}  ~~ 0.17 \leq T/n \leq 1.15 ~({\rm GeV/n}) ~~{\rm BESS}
\eeqa

The proposed sensitivites of the future AMS-02 and GAPS experiments are:
\beqa \label{sensitive}
&&\Phi_{\bar{D}} =   2.25 \times 10^{-7}  [m^2~s~sr~{\rm GeV}]^{-1}  ~~ 0.2 \leq T/n \leq 0.8 ~({\rm GeV/n})~~{\rm AMS-02}, \\
&&\Phi_{\bar{D}} =   2.25 \times 10^{-7}  [m^2~s~sr~{\rm GeV}]^{-1}  ~~ 2.2 \leq T/n \leq 4.2 ~({\rm GeV/n})~~{\rm AMS-02}, \\
&&\Phi_{\bar{D}} = 1.5 \times 10^{-7} [m^2~s~sr~{\rm GeV}]^{-1} ~~ 0.1 \leq T/n \leq 0.2 ~({\rm GeV/n})~~{\rm GAPS (LDB)} \\
&&\Phi_{\bar{D}} = 3.0 \times 10^{-8} [m^2~s~sr~{\rm GeV}]^{-1}  ~~ 0.05 \leq T/n \leq 0.25 ~({\rm GeV/n})~~{\rm GAPS (ULDB)}  \\
&&\Phi_{\bar{D}} \sim 2.6 \times 10^{-9} [m^2~s~sr~{\rm GeV}]^{-1}  ~~0.1 \leq T/n \leq 0.4 ~({\rm GeV/n}) ~~{\rm GAPS (SAT)} .
\eeqa
For higher values of antideuteron cosmic ray flux than those quoted above, antideuteron detection is expected. Since the contribution to $\Phi_{\bar{D}}$ from astrophysical background is $\Phi_{\bar{D}}^{(BG)} \sim 5 \times 10^{-9} [m^2~s~sr~GeV]^{-1}$, these experiments (with the exception of a satellite mission) will be essentially background free probes of an anomalous component of the antideuteron spectrum. Any detected signal must come from some source that contributes significantly to the low energy antideuteron spectrum. Dark matter can generate such a signal.

If dark matter annihilates, this process essentially occurs in the Earth's rest frame because the halo and Earth frame of reference are approximately the same on relativistic scales. Therefore dark matter annihilation can act as a
source of antideuterons that populates the low energy antideuteron spectrum.  If the dark matter annihilates at a large enough rate, then this signal can be detected at the AMS-02 and GAPS experiments. Furthermore, as was investigated in \cite{Baer:2005tw}, if the Dark
matter annihilates directly to colored states (e.g. $\bar{q} q$), then the final state spectrum of antideuterons
tends to be peaked at low energies due to the fact that hadronization for the final state occurs in the rest frame of the halo, but when the hadronization takes place in a boosted reference frame, for instance if the final states are boosted color singlets (e.g. $W^+W^-$), the antideuteron
spectrum tends to be peaked away from low energy. If the color singlet final states are produced at threshold however, the antideuteron spectrum will be peaked at low energies.  So, not only do antideuterons provide an essentially background free
channel in which to indirectly probe dark matter annihilations, but the detection of a signal is preferred by some annihilation modes compared to others.

It is common in many Beyond the Standard Model (BSM) physics scenarios for the dark matter to have a significant annihilation channel with hadrons in the final state. Antideuteron signals from supersymmetric neutralinos were investigated in \cite{Donato:1999gy,Baer:2005tw,Donato:2008yx}; \cite{Baer:2005tw} included an analysis of the antideuteron signal from Universal and Warped Extra dimensional dark matter candidates. In \cite{Ibarra:2009tn} the antideuteron signal was given as a function of dark matter decay rate. In \cite{Cirelli:2009uv}, heavy doublet dark matter with an enhanced present day annihilation cross section was investigated. Independent of a particular BSM physics model, the experimental reach of the GAPS experiment as a function of the present day annihilation cross section and mass for the annihilation channels of  $WW$ and $\bar{b}b$ final states was investigated in \cite{Baer:2005tw}.

The first goal of this paper is to present a model independent assessment of the dark matter mass reach of the AMS-02 and GAPS experiments for different annihilation channels. In addition to the $b\bar{b}$ and $W^+W^-$ final states, which have appeared in previous analysis, we also assess the reach for $t\bar{t}$, $hh$, and $gg$ final states as a function of dark matter mass. Furthermore, for our injection spectrum computation, we utilize the non-factorized ``coalescence scheme" recently advocated by \cite{Kadastik:2009ts}, which more accurately characterizes antideuteron production for larger kinetic energies and has significant effects even for the low energies of the GAPS and AMS-02 experiments. In Section \ref{ACRF} we describe our calculation of the antideuteron cosmic ray flux.  In Section \ref{ER} we present our results as a number of expected events as a function of dark matter mass for a given final state annihilation mode. In Section \ref{GFDM} we describe the apparent conflict between a thermal relic annihilation cross section and bounds on the spin-independent elastic scattering cross-sections derived from direct detection experiments. We then describe the basic mechanisms utilized by many models that evade this conflict.

The second goal of this paper is to understand the status of some well-motivated specific dark matter candidates considering both direct detection and indirect detection via antideuteron cosmic rays. In Section \ref{SpecMod} we apply our model independent results to a small sample of dark matter candidates, describe each model's essential features by analyzing the leading operators which govern present day annihilation and elastic scattering off of nuclei, and determine the extent to which they can be probed via antideuteron searches. Finally, in Section \ref{conc} we conclude. Appendix A contains reference material for understanding the kinematic behavior of annihilation cross-sections and elastic scattering cross-sections for general DM candidates based on operator analysis. Appendix B contains a brief description of an inelastic dark matter candidate analyzed in Section\ref{SpecMod} which is compatible with usually stringent solar neutrino bound while explaining DAMA signal.

\section{Antideuteron Cosmic Ray flux} \label{ACRF}

The antideuteron cosmic ray flux at Earth can be decomposed into primary, secondary, and tertiary components. Standard astrophysical sources produce only secondary and tertiary components; these originate from astrophysical scattering of protons off the interstellar Hydrogen and Helium and have been thoroughly studied in \cite{Chardonnet:1997dv,Duperray:2005si} and updated in \cite{Donato:2008yx}. A primary component can potentially come from the hadronic annihilation channels of dark matter. Here we focus on contributions from various types of dark matter annihilation.

Only three properties of dark matter determine its contribution to the antideuteron cosmic ray spectrum, its mass ( $m_{DM}$ ), annihilation cross section ( $\langle \sigma |v| \rangle_{ann}$ ), and fraction of annihilation into a particular final state. If there are multiple annihilation modes, their contribution just adds linearly to a good approximation. Connecting these dark matter properties to the antideuteron flux at earth requires three steps. First, one must calculate the injection spectrum of antideuterons from dark matter annihilation. Second, one must propagate the antideuterons according to a consistent model of cosmic ray propagation. Third, one must account for the effects of solar modulation. In the following subsections we review the calculations involved in these three steps.

\subsection{Injection Spectrum}
In order to calculate the injection spectrum of antideuterons from a particular dark matter model, one needs to know the dark matter mass and annihilation cross section into various final states. The dark matter mass sets the energy scale of the final state, and the composition of the final state determines how that the system hadronizes into (anti-) baryons. This process can be modeled by numerical Monte Carlo. We choose PYTHIA 6.400 \cite{Sjostrand:2006za} for our computation which were run on the Odyssey cluster supported by the FAS Research Group. Then one must make a prescription as to how a $\bar{p}$ and a $\bar{n}$ form an antideuteron. Antideuteron production, whether from $p+ H \rightarrow \bar{D} + X$ or dark matter annihilation requires such a prescription. One prescription is the so-called ``Coalescence Model" \cite{Chardonnet:1997dv}. According to this model in the $\bar{n}$ rest-frame of a $\bar{n}-\bar{p}$ system, if  the $\bar{p}$ has kinetic energy less than the
antideuteron binding energy, $B$, then one assumes the two bind to form an antideuteron. This crude model provides a good order of magnitude estimate for the formation of antideuterons. Said more precisely,  if $KE_{\bar{p}} \sim  B$ ($|k_{\bar{p}}| \sim (2m_pB)^{\frac{1}{2}}$) then antideuteron formation is assumed. So the condition that a  $\bar{p}$ and $\bar{n}$ in a hadronic final state form an antideuteron is:  $ |
\vec{k}_{\bar{n}} -\vec{k}_{\bar{p}}| < (2m_pB)^{\frac{1}{2}}\sim p_{0} \sim 70~ {\rm MeV}$. $p_0$ is referred to as the ``coalescence momentum."  We let this provide us with the conceptual understanding, but fix $p_{0}$ to reproduce the ALEPH measurement of $(5.9 \pm 1.9) \times 10^{-6}$ antideuterons per hadronic $Z$-boson decay \cite{Schael:2006fd}. Typically, when using the  ``coalescence model"  one assumes that the distribution of $\bar{n}$ and $\bar{p}$ with energy $E_{\bar{p},\bar{n}}$ is spherically symmetric and factorizes. Isospin symmetry is also usually assumed in the analysis. We will follow the analysis of \cite{Kadastik:2009ts} and compute the coalescence condition on an event by event basis. This method avoids the assumption of spherical symmetry and does not neglect phase space correlations in the production of a $\bar{n}$/ $\bar{p}$ pair.  Just as in \cite{Kadastik:2009ts}, we find that a value of $p_{0} = 160 {\rm MeV}$ is required to match the ALEPH measurement; we use this value in all calculations that follow.

\begin{figure}
\begin{tabular}{cc}
\includegraphics{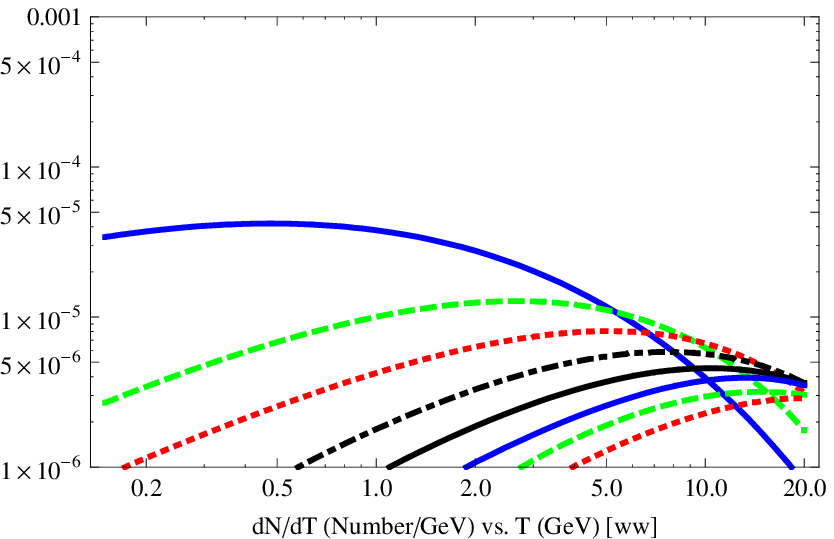} &
\includegraphics{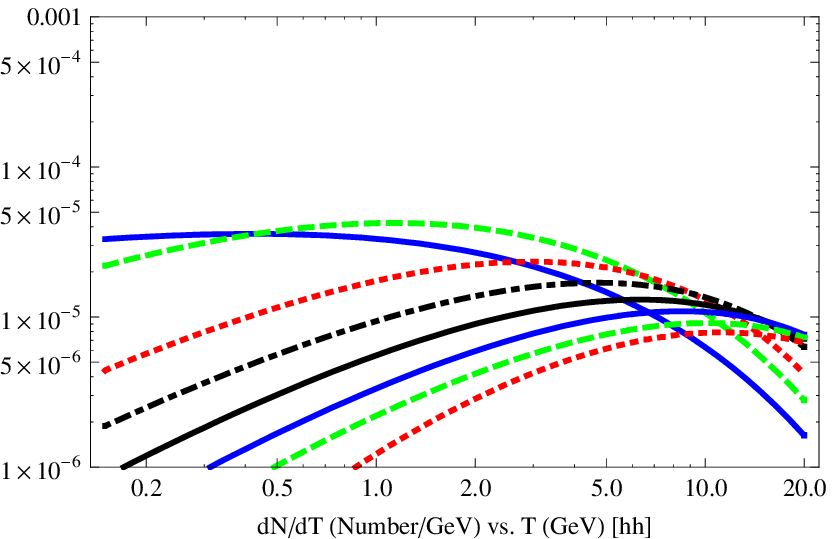} \\
\includegraphics{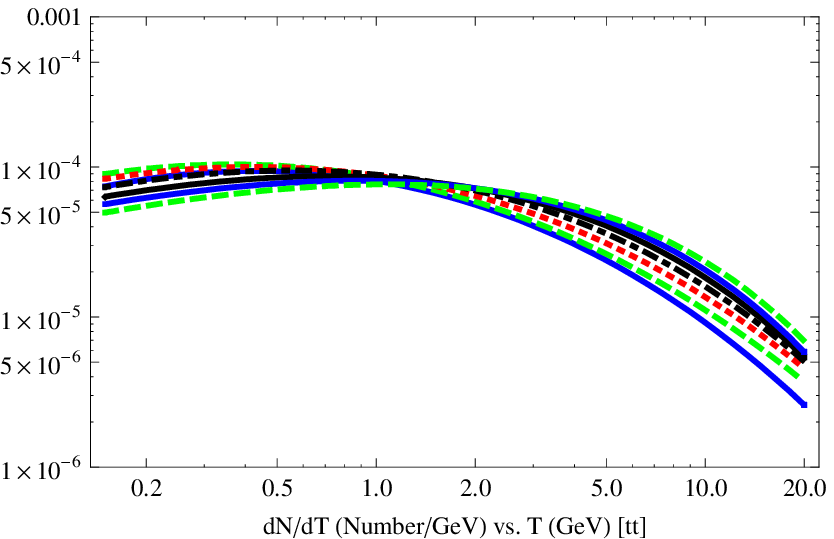} &
\includegraphics{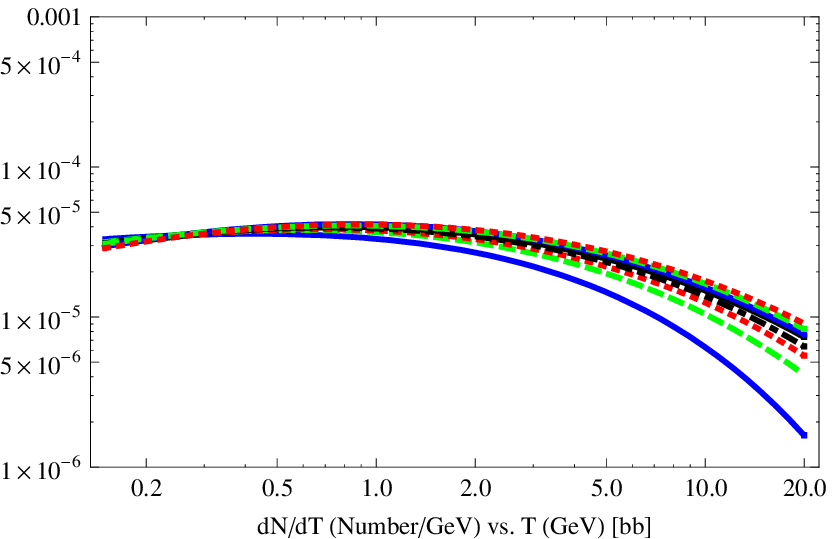} \\
\includegraphics{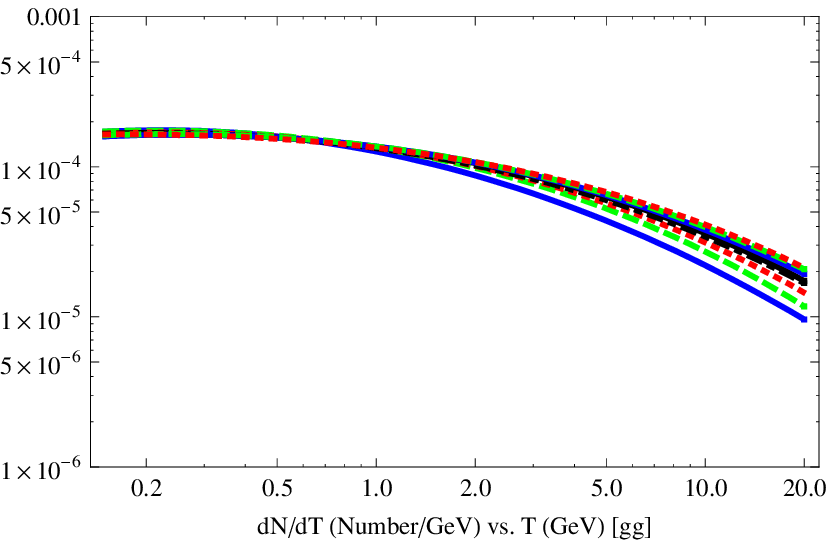}
\end{tabular}
\caption{  \label{injspect} The antideuteron injection spectrum as a function of Kinetic Energy, $T$, for DM annihilation to  $W^+W^-$, $hh$ ($m_{higgs} = 115~{\rm GeV}$), $\bar{t}t$, $b\bar{b}$, and  $gg$ final states. For each final state we plot the injection spectrum for dark matter masses of:  $m_{DM}  = 100~{\rm GeV} ({\rm blue/solid}), 200~{\rm GeV}({\rm green/dashed}), 300~{\rm GeV}({\rm red/dotted}), 400~{\rm GeV}({\rm black/dashed-dotted})$,
$ 500~{\rm GeV}({\rm black/solid}), 600~{\rm GeV}({\rm blue/solid}), 700~{\rm GeV}({\rm green/dashed}), 800~{\rm GeV}$({\rm red/dotted}) as computed using PYTHIA 6.400 and a coalescence momentum $p_0= 160 ~ {\rm MeV}$. However, for $hh$ and $t\bar{t}$ final states the lowest dark matter mass plotted is $m_{DM} = 150~{\rm GeV}({\rm blue/solid})$ and $m_{DM} = 200~{\rm GeV}({\rm blue/solid})$, respectively.}
\end{figure}

We have computed the injection spectra for antideuterons for specific final states.  In Fig. \ref{injspect} we have plotted the injection spectrum for antideuterons for $\bar{t}t$, $\bar{b}b$, $h^0h^0$,  $gg$, and $W^+W^-$ final state annihilation channels for various masses.  One can see a clear distinction between colored and color neutral final states for larger values of the dark matter mass.  The colored final states yield a spectrum that remains peaked at low energies as the mass increases. However the color neutral final states yield spectra that are peaked at higher energies as the mass increases. This is due to the fact that hadronization takes place in a boosted reference frame for energetic color neutral final states. But notice that when the color neutral states are produced near threshold, the fact that they are nearly at rest implies hadronization takes place in the lab frame resulting in a spectrum not too different than the colored final states. Note also that the $t\bar{t}$ spectrum is generally larger than the $b\bar{b}$ spectrum for the same dark matter mass because $t \rightarrow b + W^+$ before hadronizing. Then, in addition to the $b\bar{b}$ component of the $t\bar{t}$ final state, when the $W^+$ decays, and $2/3$ of its decay products are hadrons, the total final state will contain more hadrons, and thus more antideuterons, in general. A related feature of the top injection spectrum is that it generally contains a sightly higher energy antideuteron component due to the fact that as the dark matter mass increases, the Ws carry more of the initial top's kinetic energy and tend to populate the higher energy component of the antideuteron spectrum. A similar disparity exist between then $W^+W^-$ and $h^0h^0$ spectra but here it is simply that for a light higgs $h^0 \rightarrow b\bar{b}$ $100\%$  while $W^+ \rightarrow hadrons $ only $2/3$ of the time. So, for a given dark matter mass the light higgs final states have a higher antideuteron yield than do the Ws. There is also a small effect from the fact that the higgs are heavier than the W's and thus are never quite as boosted as the Ws. The distinction between the two spectra becomes more dramatic when the higgs mass is significantly larger than the W-mass. For heavier higgs ($m_{higgs} \geq 140~{\rm GeV}$), it decays dominantly to $WW$ or $ZZ$ and one finds that the spectrum resembles the $W$ final state spectrum as the mass increases.  In order to simplify our analysis, in Fig. \ref{injspect} as well as for the remainder of the paper, we fix $m_{higgs} = 115 ~ {\rm GeV}$. We have not given explicit injection spectra for $ZZ$ and light quark final states because these are very similar in shape and magnitude to the $W^+W^-$ and $b\bar{b}$ injection spectra, respectively.

Finally, we comment that the shape of these injection spectra are flatter compared to the injection spectra one derives from using the factorized ``coalescence model" due to the fact that higher energy antideuteron production is not suppressed using an event by event calculation \cite{Kadastik:2009ts}.

\subsection{Dark Matter Profile}

The shape of the dark matter halo profile can be parameterized as an NFW \cite{Navarro:1996gj}, Einasto \cite{Graham:2005xx}, or Isothermal \cite{Bahcall:1980fb} profile. We will restrict our attention to the Einasto profile due to the fact that it provides the most conservative signal while still being preferred by numerical simulations.

\beq
\rho_{Ein}(r) = \rho_{\odot}\exp{\left[-2\left[\left(\frac{r}{r_s}\right)^{\alpha}-\left(\frac{r_{\odot}}{r_s}\right)^{\alpha}\right]/\alpha \right]}
\eeq
with $\rho_{\odot} = 0.4 ~{\rm GeV}/{\rm cm}^3$, $r_s = 20 ~ {\rm kpc}$, and $r_{\odot} = 8.5 ~{\rm kpc}$.

\subsection{Propagation}

In order to connect the injection of antideuterons in the galactic halo to the flux of antideuterons near the Sun, one must
propagate the antideuterons in the galactic halo. For propagation, our computations use a two dimensional diffusion model and our discussion in this section follows the notation of \cite{Maurin:2001sj,Barrau:2001ev,Donato:2001ms}(for reviews see \cite{Ginzburg:1976dj,Strong:2007nh}). In order to parameterize the uncertainty in propagation, we compute CR propagation with MIN, MED, and MAX  propagation parameters as listed in Table \ref{models} and given in \cite{Donato:2003xg}.

\begin{table}[t!]
\centering
\begin{tabular}{|c||c|c|c|c|} \hline
Model & \quad $\delta$ \quad &\quad $K_0$  (${\rm kpc}^2/{\rm Myr}$) \quad & \quad $L$(kpc) \quad &\quad $V_C$  (km/s) \quad \\ \hline
MIN  & \quad 0.85 \quad & \quad 0.0016 \quad & $1$ & 13.5\\
MED   & 0.70 & 0.0112 & $4$& 12  \\
MAX   & 0.46 & 0.0765 & $15$& 5 \\ \hline
\end{tabular}
\caption{\label{models} Propagation models}
\end{table}

The diffusion equation for charged cosmic rays, neglecting energy losses, is given by
\beqa
\frac{d}{dt}\psi(r,z,E) &=& Q(r,z,E) - 2h\delta(z)\Gamma_{ann}(E)(n_H +4^{\frac{2}{3}} n_{He}) \psi(r,z,E) \nonumber \\ &+& K(E)\left( \frac{\partial^2}{\partial z^2} + \frac{1}{r}\frac{\partial}{\partial r}r\frac{\partial}{\partial r} \right)\psi(r,z,E) -V_C\frac{\partial}{\partial z} \psi(r,z,E)
\eeqa
where $\psi(r,z,E)$ is the number density of a cosmic ray species.  Steady state solutions are given by setting $\frac{\partial}{\partial t} \psi(r,z,E) = 0$. Also, the region over which this equation is solved is assumed to be a cylinder with height L, which can vary as in Table \ref{models}, and a radius fixed at $R = 20 {\rm kpc}$. A conservative estimate of the CR flux can be made by assuming that $\psi(R,L,E)=0$, at the boundary. A second boundary condition is given by assuming continuity through  the $z=0$ plane. Solutions are expressed in terms of a Bessel Series Expansion.

It is common to write fluxes in terms of particle kinetic energy ($T$) rather than energy ($E$), so we do this in what follows. The function $Q(r,z,T)$ depends on the injection spectrum as
\beq
Q(r,z,T) =\frac{1}{2} \left< \sigma v \right> \left( \frac{ \rho(r,z)}{m_{DM}} \right)^2 \frac{dN}{dT}.
\eeq
The Hydrogen and Helium number density in the galactic disk are: ${\rm n_H} = 1/{\rm cm}^3$ and ${\rm n_{He} }= .07~{\rm n_H}$, respectively. $\Gamma_{ann}(T)$ describes the depletion of the antideuteron population due to dissociation after scattering from interstellar Hydrogen and Helium\footnote{Note that $\Gamma$, as it is used here, does not stand for a decay width.}, $\Gamma_{ann}(T) =  \sigma(T)v$ where
\beq
\sigma(T)=\{ \begin{array}{cc} 661(1+0.0115T^{-0.774}-0.984T^{0.0151}){\rm mb} & T < 15.5~{\rm GeV}\\
36T^{-0.5} {\rm mb},  & T \geq 15.5 ~{\rm GeV} \end{array}
\eeq
\beq
K(T) = K_0 (R(T))^\delta, ~~ R(T) = \frac{(E(T)^2-m^2)^{\frac{1}{2}}}{|Ze|} ,~~~ E(T) = T+m
\eeq
with T in units of GeV. $K_0$ sets the overall scale for charged particle diffusion and $V_C$ accounts for a ``galactic wind" that pulls charged particles out of the disk. $K_0$ and $V_C$ for different sets of propagation parameters are listed in Table I.
The solution for the number density of CR's is given by:
\beq
\psi(R_{\bigodot},0,T) = \sum_i^{} N_i(0,T)J_0(\frac{R_{\bigodot}}{R}\xi_i)
\eeq
where
\beq
N_i(0,T)=e^{\frac{-V_CL}{2K(T)}} \times \frac{y_i(L)}{A_i\sinh{(S_iL/2)}}
\eeq
with
\beq
y_i(T) = 2\int_0^Le^{(\frac{V_c}{2K(T)}(L-z'))} \times \sinh{(\frac{S_i}{2}(L-z'))}q_i(z')dz',
\eeq
\beq
q_i(z)= \frac{2}{ J^2_1(\xi_i)}\int_0^1\rho Q(\rho,z)J_0(\xi_i\rho)d\rho, ~~ \rho = \frac{r}{R},
\eeq
\beq
S_i = \left( \frac{V_C^2}{K^2(T)} + 4 \frac{\xi^2_i}{R^2}\right)^{\frac{1}{2}}, ~~ A_i = 2h\Gamma_{ann} + V_C + K(T)\coth{(S_i L/2)}
\eeq
where $\xi_i$ are the zeros of $J_0$.
The cosmic ray flux at the Sun is given by
\beq
\Phi(R_{\bigodot}, 0, T) = \frac{v(T)}{4\pi}\psi(R_{\bigodot}, 0, T).
\eeq

\subsection{Solar Modulation}

Finally, to relate the flux at the Sun's position to the flux at Earth, we must take into account effect of solar modulation. Solar modulation reduces the flux of low energy cosmic ray species. This final step is of particular importance since we are interested in the low-energy part of the antideuteron spectrum, where solar modulation effects are most important. We take into account solar modulation using the Gleeson Axford force field model \cite{mod}:
\beq
\Phi_{\bigoplus}(T_{\bigoplus}) = \frac{2mT_{\bigoplus} + T_{\bigoplus}^2}{2mT + T^2}\Phi(T), ~~~ T = T_{\bigoplus} + e\phi_F.
\eeq
We will take $\phi_F = 500{\rm MV}$ in what follows.

\subsection{Backgrounds}

When considering whether or not a predicted signal from a primary source is large enough to be significant or not,  it is important to understand the astrophysical sources which generate the background. The background computation of secondary and tertiary antideuterons due to protons colliding with interstellar Hydrogen and Helium were originally carried out in \cite{Chardonnet:1997dv,Duperray:2005si}. For the analysis that follows, we use the expected antideuteron flux at the top of the atmosphere for ``MED" propagation parameters as computed in \cite{Donato:2008yx}.

\section{Experimental Reach for Certain Final States} \label{ER}

We now propagate the various injection spectra according to the procedure outlined above, assess the sensitivity of antideuteron search experiments, and comment on uncertainties in the results.

\subsection{Results}

\begin{figure}
\begin{tabular}{cc}
\includegraphics{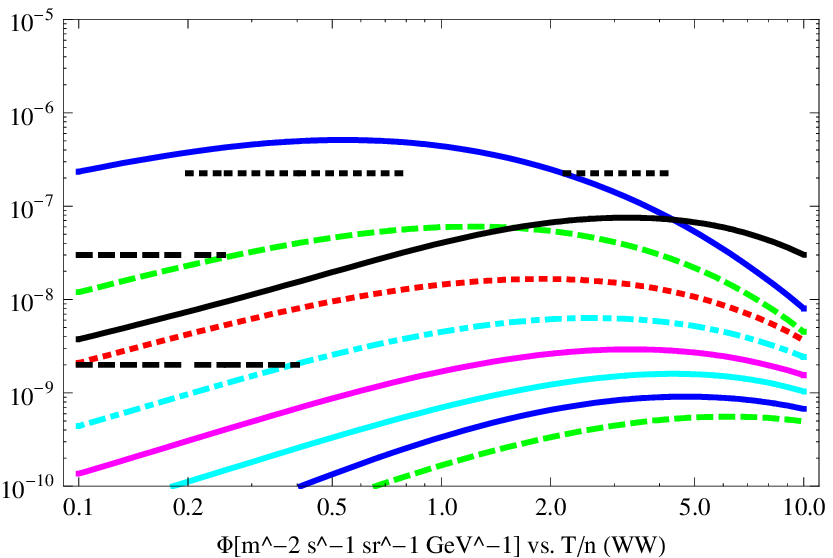} &
\includegraphics{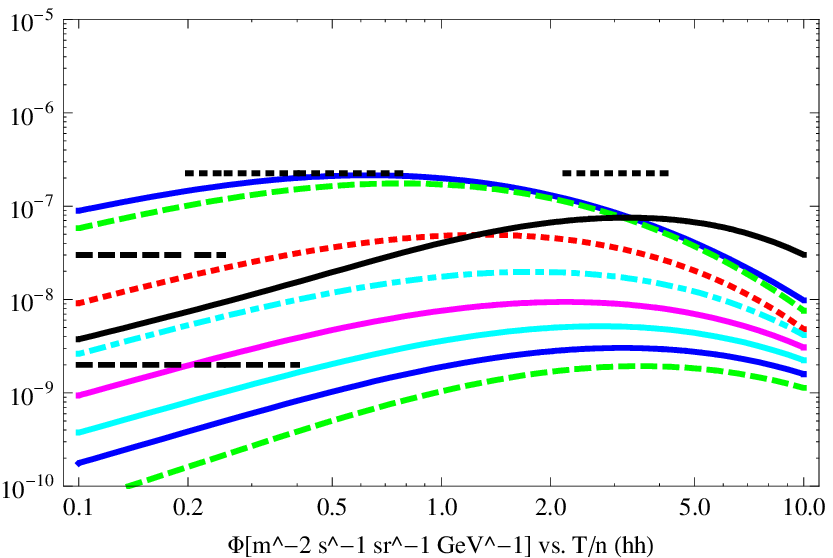} \\
\includegraphics{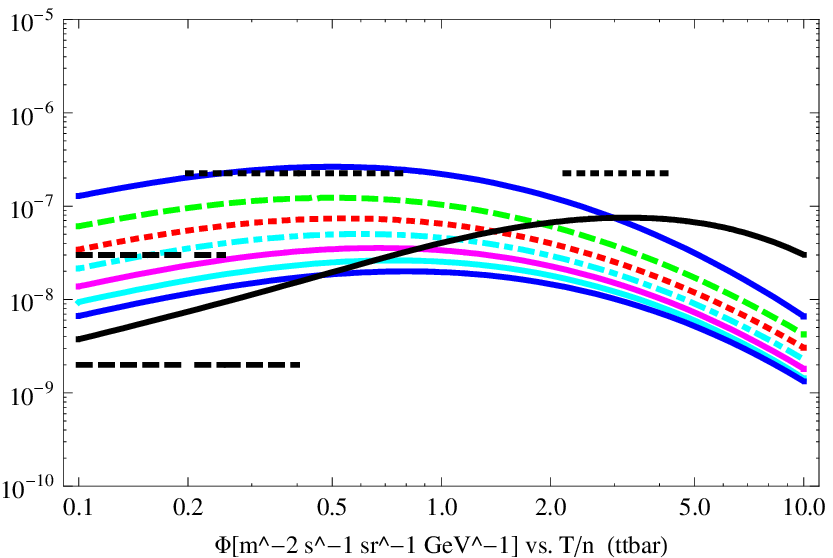} &
\includegraphics{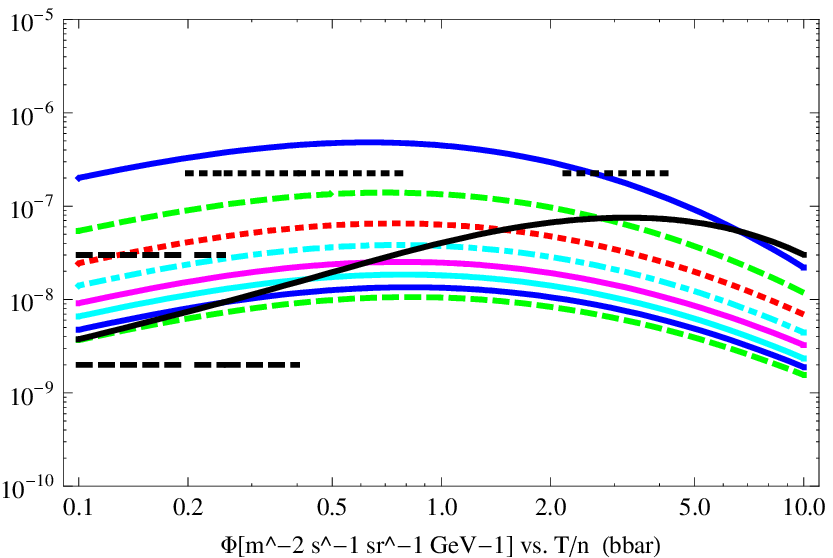} \\
\includegraphics{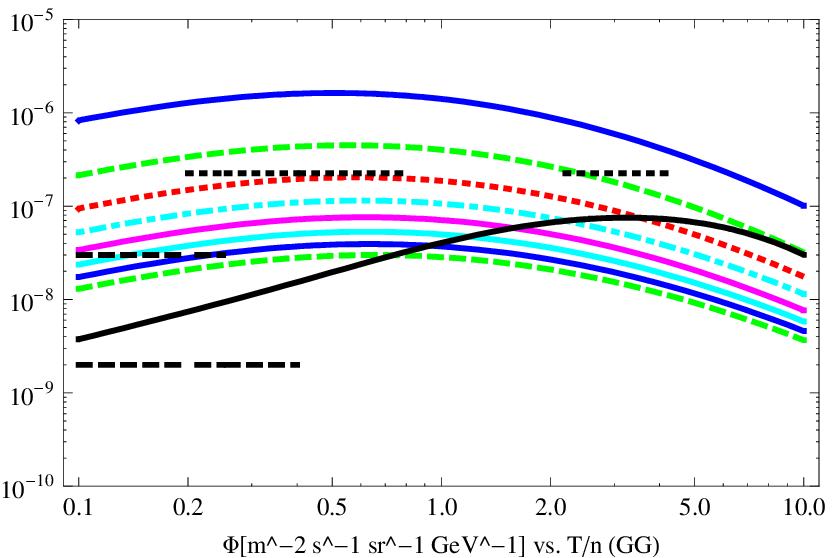} \\
\end{tabular}
\caption{The antideuteron reach of the AMS-02 (black dotted) and GAPS (ULDB) (black upper-Dashed) and GAPS (SAT) (black lower-dashed) experiment for dark matter annihilation to $W^+W^-$, $hh$ ($m_{higgs} = 115~{\rm GeV}$), $\bar{t}t$, $b\bar{b}$, and  $gg$ final states. In each case the dark matter present day annihilation is set to $\langle \sigma |v| \rangle = 1~{\rm pb}$. The fluxes are plotted for different dark matter masses from $m_{DM} = 100 -800 ~{\rm GeV}$ as in Fig. \ref{injspect}. The flux decreases as the mass increases. Also, the astrophysical background is given by the solid line. Propagation is done with ``MED" parameters.}
 \label{ADflux}
\end{figure}

Modulated antideuteron fluxes and experimental sensitivities are presented in \fig{ADflux}. The fluxes are plotted for a fixed dark matter annihilation cross-section of $\langle \sigma |v| \rangle_{ann} = 1 ~{\rm pb}$ and various dark matter masses and annihilation modes and ``MED" propagation parameters. The predicted background antideuteron flux as calculated in \cite{Donato:2008yx} is also plotted. We shall first discuss the features of the flux originating from dark matter annihilations (signal) and then consider the effects of the background when we estimate the experimental reach of the AMS-02 and GAPS experiments.

\begin{figure}
\begin{tabular}{cc}
\includegraphics{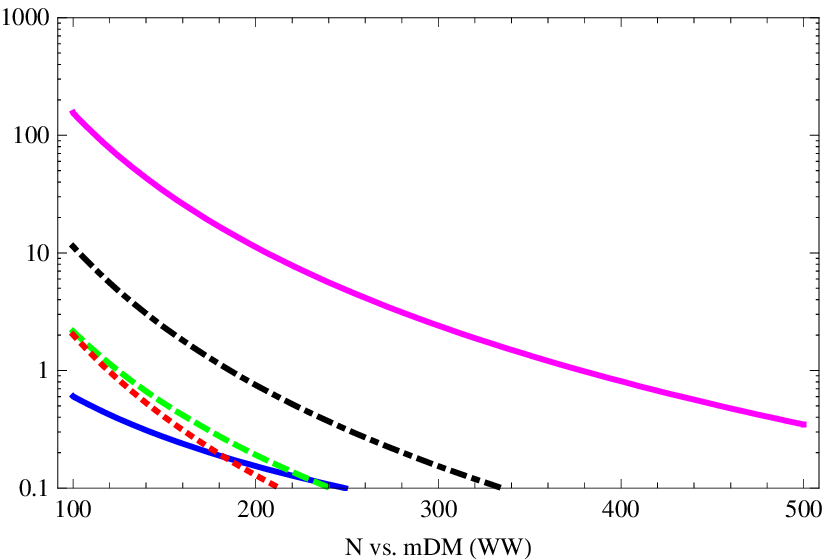} &
\includegraphics{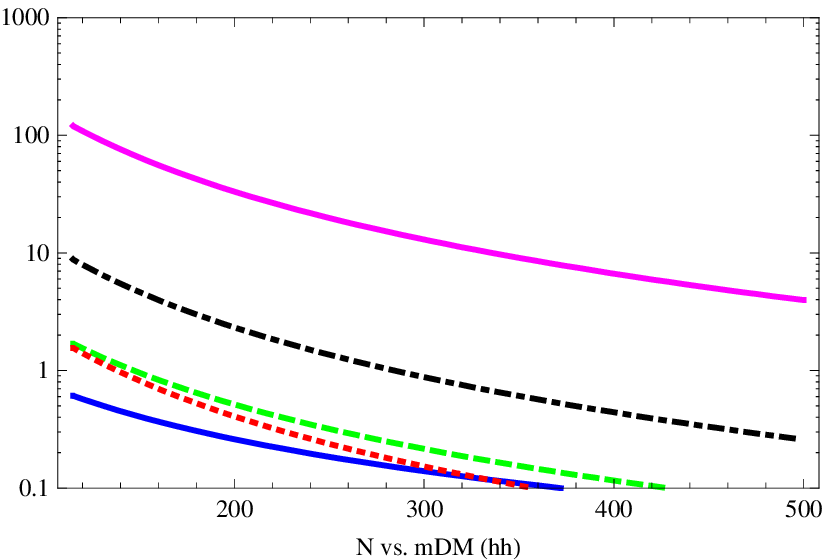} \\
\includegraphics{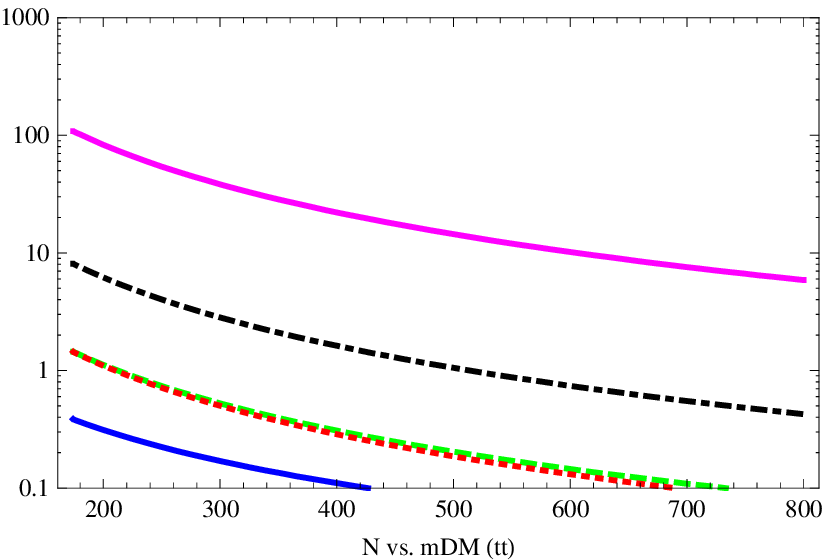} &
\includegraphics{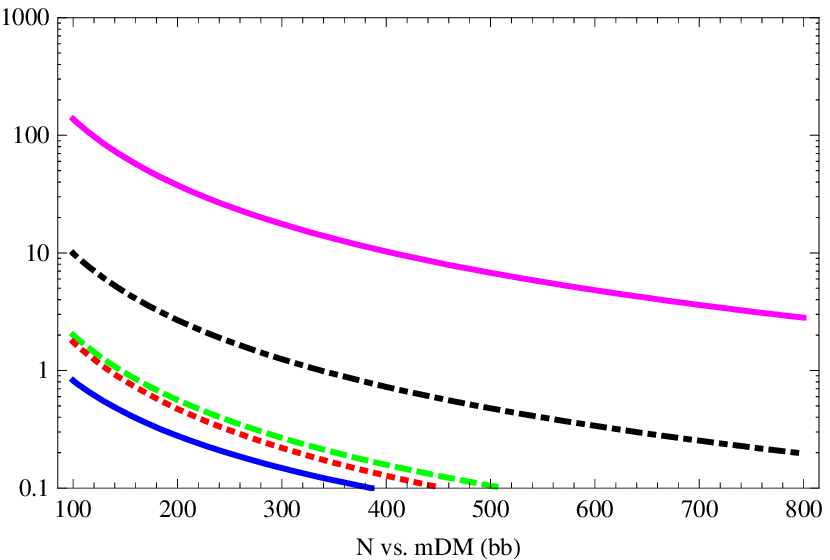} \\
\includegraphics{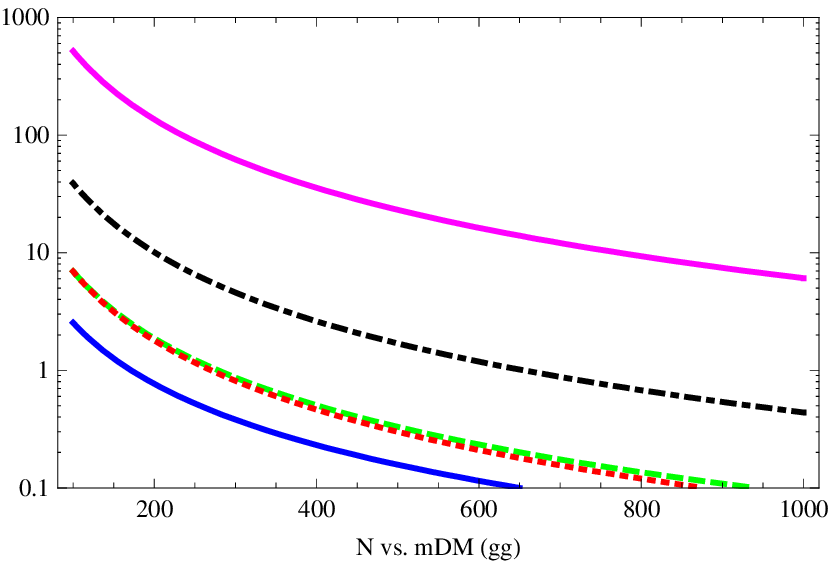} \\
\end{tabular}
\caption{The expected number of primary antideuterons for the AMS-02 (high) (blue/solid), AMS-02 (low) (green/dashed, )GAPS (LDB) (red/dotted), GAPS (ULDB) (black/dot-dashed) and GAPS (SAT) (pink/solid/upper line) experiments for dark matter annihilation to $W^+W^-$, $hh$ ($m_{higgs} = 115~{\rm GeV}$), $\bar{t}t$, $b\bar{b}$, and  $gg$ final states. $\langle  \sigma |v| \rangle_{ann}= 1 ~{\rm pb}$. Propagation is done with ``MED" parameters.}
  \label{Nplot}
\end{figure}

	Here we see that for each final state there exists some mass range for which AMS-02 and GAPS are sensitive to that particular dark matter annihilation mode. Let us emphasize that for even the largest antideuteron signal listed here, the antiproton spectrum is predicted to remain consistent with current BESS and PAMELA data, this is true even if one uses the ``MAX" set of propagation parameters for antiprotons. Assuming that the sensitivity lines in \fig{ADflux} correspond to the fluxes required for a detection of one antideuteron, we convert each flux in \fig{ADflux} into a number of expected antideuteron events from primary sources ($N^{prim}_{\bar{D}}$) at each of the experiments (AMS-02 (high), AMS-02 (low), GAPS (LDB), GAPS (ULDB), GAPS (SAT)) and plot the results in \fig{Nplot}. From \fig{Nplot}, it is clear that  the antideuteron signal for dark matter annihilation to colored final states does not fall off as rapidly as for color neutral states. This reflects the fact that hadronization and fragmentation takes place in a boosted frame of reference for higher energy color neutral final states.  From this plot we can also see that the predicted number of antideuteron events can be 10 for some scenarios and even 100 for a GAPS level sensitivity on a satellite. However, for larger masses the number falls quite rapidly.

Notice, again, that the shape of the flux differs from the flux one would derive using the factorized ``coalescence model" due to the fact that the injection spectra predicted from the two models differs for larger antideuteron kinetic energies. In this case the flux is sustained at higher energies. This is important when considering experiments like AMS-02 which are sensitive to antideuterons with higher kinetic energies.

\begin{table}[t!]
\centering
\begin{tabular}{|c||c|c|c|c|c|c|c|} \hline
Experiment &  \quad $\bar{q}q$ \quad&  \quad $\bar{t}t$ \quad & \quad $h^0h^0$ \quad &\quad $gg$ \quad&\quad $W^+W^-$ \quad & $N_{crit}$ \\ \hline

AMS-02 high (3$\sigma$)& 50 & $< ~ m_t$  & $< m_h$  &  100  &  $<m_W$ & 3  \\
AMS-02 low (3$\sigma$)& 100 & $<m_t$  &$<m_h$  &  200  &  100 & 2  \\
GAPS (LDB) (3$\sigma$)& 140 & 200  & 140 & 300 & 120 & 1\\
GAPS (ULDB) (3$\sigma$)& 250 & 400 & 250 & 500 & 160 & 2 \\
GAPS (SAT)  (3$\sigma$)& 500 &  700& 500 & 900 & 240  & 10\\ \hline
AMS-02 high (5$\sigma$)& $30 $ &   $~< m_t~ $ &  $< m_h$  &  $60$ & $< m_W$ & 6\\
AMS-02 low (5$\sigma$)& $70 $ &  $~< m_t~ $ &  $< m_h$  &  140  & $< m_W$ & 4\\
GAPS (LDB) (5$\sigma$)& $ 75 $&  $~< m_t~ $ & $< m_h$ & 150 & $< m_W$  & 3\\
GAPS (ULDB) (5$\sigma$)& 150  & 220 & 150 & 300 & 120 & 5\\
GAPS (SAT)  (5$\sigma$)& 360 & 550 & 300 & 670 & 200 & 16\\ \hline
\end{tabular}
\caption{\label{reach} Mass reach for various experiments (all masses in GeV units). Masses quoted denote the dark matter mass for which the signal + background flux predicts a number of events equal to $N_{crit}$. See text for a definition of $N_{crit}$. For each annihilation mode, $100\%$ branching, $\langle \sigma |v| \rangle_{ann} = 1~{\rm pb}$, MED propagation parameters, and coalescence momentum of $p_0= 160 ~{\rm MeV}$ is assumed. }
\end{table}

We parameterize the background as recently given in \cite{Donato:2008yx} which builds on the work of \cite{Chardonnet:1997dv,Duperray:2005si} and have inferred from their results and the quoted experimental sensitivities, \eq{sensitive}, that the expected number of background events at each of the experiments is:
\beqa
&& ~~~~~~~~~~{\rm AMS-02 (high)}:~ b= 0.3,~~~~~~~~~~~~ ~~~ {\rm AMS-02 (low)}:~b= 0.1, ~~~ \nonumber \\
&& {\rm GAPS (LDB)}:~b=0.04, ~~~~~{\rm GAPS (ULDB)}:~b=0.2, ~~~~~ {\rm GAPS (SAT)}:~b=3.5
 \eeqa

 In order to characterize the mass reach of antideuteron search experiments we use a statistical analysis similar to that of \cite{Baer:2005tw}, where we assume a Poisson distribution for the detection of antideuteron events. In Table \ref{reach} we present, for various annihilation modes, the dark matter mass for which the number of expected antideuteron events equals $N_{crit}$, where $N_{crit}$ is determined as follows. $N_{crit}$ is the lowest value of $N$ satisfying the following inequalities:
\beq
\sum_{n=0}^{N-1}P(n,b) > 0.997  ~~~~~ (3\sigma), ~{\rm or}~~~~\sum_{n=0}^{N-1}P(n,b) > 0.9999994  ~~~~~ (5\sigma)
\eeq
where $P(N,b) = (b^Ne^{-b})/N!$ and $b$ is the expected number of background events. In other words the probability of detecting the number of antideuterons listed in the $N_{crit}$ column of Table \ref{reach} for each experiment is:
\beq
P_{3\sigma} (N \geq N_{crit}) < .003, ~~~~~{\rm or} ~~~~~~~P_{5\sigma} (N \geq N_{crit}) < 0.0000006
\eeq
In these cases a detection of $N_{crit}$ antideuteron events implies that one is seeing an exotic contribution to antideuteron cosmic rays at a confidence level $ > 95\%$ and $ > 99.9999\%$, respectively. The masses in Table \ref{reach} are, given an annihilation mode, the largest mass for which the antideuteron flux arising from the sum of the dark matter component and the background yields an expected number of events equal to $N_{crit}$.  The number of primary events expected from a dark matter source can be inferred from Fig. \ref{Nplot}. This means that if dark matter is annihilating to $t\bar{t}$ final states with a mass $m_{DM} < 400 ~ {\rm GeV}$ or $hh$ with $m_{DM} < 250 ~ {\rm GeV}$  then one should expect to see a significant source of primary antideuterons at GAPS (ULDB). Compared to the fact that the antiproton measurements bound dark matter masses of $50~{\rm GeV}$ \cite{Baer:2005tw} with the same annihilation channel, GAPS(ULDB) will be able to probe dark matter masses which are greater by an order of magnitude.

\subsection{Uncertainties} \label{unc}

Theoretical uncertainties arise at various points in the computation that produced the results in \fig{ADflux} and \fig{Nplot}. They are uncertainty in the coalescence momentum ($p_0$) that determines the probability that an antideuteron forms during hadronization, the cosmic ray propagation parameters, the shape of the dark matter halo, and solar modulation.  A discussion of these uncertainties can also be found in \cite{Baer:2005tw} and \cite{Donato:2008yx}.

The predicted antideuteron flux is most sensitive to the choice of  ``coalescence momentum" ($p_o$) used to calculate the injection spectrum of antideuterons from dark matter annihilation. Because the ``coalescence model"  keeps only events within a 3-d volume of phase space set by the parameter $p_0$, the overall flux of antideuterons depends on $p_0$ as: $\Phi_{\bar{D}} \sim p_0^3$. The choice and uncertainty of $p_0$ is therefore very important. Recently,  \cite{Kadastik:2009ts} has argued an event by event computation of the coalescence condition is essential for a proper computation of higher energy injection spectra. After matching to ALEPH's measurement of $(5.9 \pm 1.9) \times 10^{-6}$ antideuterons per hadronic $Z$-boson decay \cite{Schael:2006fd}, the resulting  $p_0 =160 \pm 17 ~{\rm MeV}$, translating into an uncertainty in the flux of $90\%$.  This is the uncertainty estimate for simulating dark matter annihilation for a dark matter mass of $45~{\rm GeV}$. Based on the similar values for $p_0$ deduced at lower energies \cite{Antipov:1971zs,Duperray:2005si} compared to those inferred from the ALEPH data, we expect that this uncertaintly in $p_0$ should not increase significantly as the center of mass energy of the collision increases to $1 ~{\rm TeV}$.

The next significant uncertainty in the antideuteron flux comes from the uncertainty in the parameters of cosmic ray propagation listed in Table \ref{models}. Roughly, these amount to a variation in the flux by an order magnitude above and below the allowed flux predicted from using ``MED" propagation parameters.

The dark matter halo profile is also uncertain with NFW (Navarro-Frenck-White) , Einasto, and Isothermal profiles as standard choices. When the dark matter density, $\rho$, becomes large, dark matter annihilation is enhanced as $\rho^2$. Since the flux of antideuterons is sensitive to the halo shape at the center of the galaxy, the flux can depend strongly on the choice of halo profile. In \cite{Donato:2008yx}, it was shown that there is a correlation between the uncertainty due to the halo profile and the choice of propagation model.  For MED and MIN propagation parameters the uncertainty in the low energy region of antideuteron flux is $\mathcal{O}(5\%)$ or less, while for MAX propagation parameters this uncertainty can be as large as $\mathcal{O}(50\%)$.

Overall the uncertainty in the theoretical prediction for the comic ray antideuteron flux is dominated by the uncertainties of cosmic ray propagation. The actual flux could be a factor of 10 larger or smaller than the average values used in this paper.

\section{General Bounds/Features of Thermal Dark Matter} \label{GFDM}

In the previous section, we have essentially done a `model-blind' study of antideuterons as a probe of thermal WIMP DM because we have assumed $\langle \sigma |v| \rangle_{ann} = 1~{\rm pb}$ and varied the DM annihilation mode and mass. In the remainder of this work we will apply this general analysis to specific DM candidates. Much existing work on antideuteron DM searches focus on the reach for particular, well-motivated models like SUSY LSP \cite{Donato:1999gy,Baer:2005tw,Donato:2008yx};  we shall consider these and other models in the following sections.

Before addressing specific models, in this section we will start with a very broad view and examine the conditions under which general WIMP DM can simultaneously yield the right thermal relic density and satisfy current direct detection (DiDt) bounds. See \cite{Kurylov:2003ra,Beltran:2008xg,Cao:2009uv} for related discussions. The relevant DM properties are the DM quantum numbers (e.g. charge and spin) and the type of interaction it has with the Standard Model. DM properties can be organized by considering general DM bilinear operators and the Standard Model bilinear operators to which DM couples. We leave the detailed discussion and summary of these operator properties to Appendix A and will reference the results listed there as needed during the following sections. The important point that one gains from such an analysis is that a variety of thermal WIMP DM models can have very suppressed DiDt rate while still providing a good signal for indirect detection (IdDt) via antideuterons. The general discussion here and in Appendix A also provide a more systematic way of understanding the specific models we will consider in the next section.\\

In general, there are four important cross-sections for a thermal dark matter candidate. The first is its annihilation cross section at early times $\langle \sigma |v| \rangle_{therm}$. This determines the thermal relic abundance of dark matter. For this work we are assuming that  $\langle \sigma |v| \rangle_{therm} = 1~{\rm pb}$ which leads to $\Omega_{DM}h^2=0.12$. The second is the present day annihilation cross-section $\langle \sigma |v| \rangle_{ann}$. This is important for indirect detection. The third and fourth are spin-independent (SI) and spin-dependent (SD) scattering cross section off nucleons: $\sigma_{SI}$ and $\sigma_{SD}$, respectively.  These are important for direct detection.

In this section, we first discuss the general correlation between $\langle \sigma |v| \rangle_{therm}$ and $\sigma_{SI}$ and how current constraints on $\sigma_{SI}$ from Direct Detection (DiDt) experiments imply the inequality:
\beq \label{therm/SI}
{\langle \sigma |v| \rangle_{therm} \over \sigma_{SI} }\geq 10^{7},
\eeq
for a thermal relic of EW scale mass.  First, we elaborate on why a thermal relic must satisfy this inequality. Then we discuss the essential features allowing models to satisfy this inequality.   In particular we discuss how the physics determining  $\langle \sigma |v| \rangle_{therm}$ can be significantly different than $ \sigma_{SI} $.

  \subsection{Thermal Relic Density versus DiDt Bounds}

Assuming that $\Omega_{DM}h^2=0.12$ arises because dark matter is a thermal relic implies that $\langle \sigma|v| \rangle_{therm}= 1~\rm pb $ at freezeout. The  CDMSII \cite{Ahmed:2009zw} and XENON100 \cite{XE100} DiDt experiments bound the spin-independent (SI) scattering cross section of DM on target nucleon of roughly $\sigma_{SI}\lesssim10^{-7}~{\rm pb}$ for a typical WIMP in the mass range $m_{\chi}\sim10-10^3~{\rm GeV}$. In general these two cross sections can be correlated via crossing symmetry of the Feynman diagram that controls dark matter annihilation. Consider, for example, a scalar dark matter candidate ($\chi$ and its antiparticle) which annihilates to quarks, leptons, W/Z with `unbiased' universal couplings, has DM mass $m_{\chi}$, mediator mass $M$, width $\Gamma_M$, and mediator couplings to dark matter and Standard Model state $g_1, g_2$ (for scalar DM $g_1$ is of mass dimension 1), respectively. Then we get an effective Fermi coupling for the related operator $\chi^{\dagger}\chi\bar{q}q$,
  \beq
  G=\frac{g_1g_2}{[(4m_{\chi}^2-M^2)^2+\Gamma_M^2M^2]^{1/2}}.
  \eeq
 Let the total annihilation cross section be $\langle \sigma|v| \rangle_{therm}=1~{\rm pb}$. In order to relate this to DiDt, we focus on processes involving the up quark. With a color factor and universal couplings, it is reasonable to assume that the annihilation fraction to u quarks is $10\%$.
  \beq
  \langle \sigma|v| \rangle_{therm}^u=\frac{N_c\overline{|M|^2}}{8\pi(2m_{\chi})^2}
  \eeq
  where after spin sum, $\overline{|M|^2}\approx8G^2m_{\chi}^2$. So,
  \beq \label{ann}
\langle \sigma|v| \rangle_{therm}^u=\frac{3(g_1g_2)^2}{4\pi[(4m_{\chi}^2-M^2)^2+\Gamma_M^2M^2]}=10^{-37}\rm cm^2.
  \eeq
  Here we get a required relation
  \beq
  \frac{(g_1g_2)^2}{4\pi[(4m_{\chi}^2-M^2)^2+\Gamma_M^2M^2]}\approx3\times10^{-38}\rm cm^2\label{ann}.
  \eeq
 Crossing the Feynman diagram we get an associated process for DiDt (here we need to sum over all light quarks and the heavy quark contribution via a gluon loop). Take scattering off the proton as an example. The cross section is \footnote{The quantity DiDt experiments bound is the DM-nucleus scattering cross section per nucleon, which can be exactly compared to $\sigma_{\chi p}$ computed here if DM equally couples to p and n. This is true for the example presented here where DM has universal scalar coupling to all quarks. There can be $O(1)$ difference if DM couples to neutrons and protons differently, but this does not alter the main point here.}
  \beqa \label{mp/mdm}
 \sigma_{\chi p}&=&\frac{1}{4\pi}\frac{m_p^2}{(m_{\chi}+m_p)^2}\frac{(g_1g_2)^2}{M^4}
  \left(\sum_{q=u,d,s}\frac{m_p}{m_q}f^p_{Tq}+\sum_{q=c,b,t}\frac{m_p}{m_q}\frac{2}{27}f^p_{TG}\right)^2\label{DD0}
  \\\nonumber&\approx&\frac{1}{\pi}\frac{m_p^2}{m_\chi^2}\frac{(g_1g_2)^2}{M^4}
    \eeqa
  where $f^p_{TG}, f^p_{Tq}$ are proportional to gluon and quark matrix element in the nucleon. To get the second line we have used: $m_{\chi}\gg m_p$, with the measured values of: $f^p_{TG}, f^p_{Tq}$: $f^p_{Tu}=0.020\pm0.004, f^p_{Td}=0.026\pm0.005, f^p_{Ts}=0.118\pm0.062, f^p_{TG}=1-\sum_{u,d,s}f^p_{Tq}$ which is about $0.84$.\\
  Meanwhile the current bounds from XENON100 and CDMS-II imply for $m_\chi\sim100\rm GeV$
\beq
  \frac{1}{\pi}\frac{m_p^2}{m_\chi^2}\frac{(g_1g_2)^2}{M^4}\lesssim3\times10^{-44}\rm cm^2\label{DD}.
\eeq
However, thermal relic density requirements (eq.(\ref{ann})) imply
\beq
\sigma_{SI}=\sigma_{\chi p}=\frac{4m_p^2}{m_{\chi}^2}\frac{(4m_{\chi}^2-M^2)^2+\Gamma_M^2M^2}{M^4}3\times10^{-38}\rm cm^2\label{DD}.
\eeq
For EW scale $m_{\chi}$ and $M$, this implies $\sigma_{\chi p}\sim10^{-41}\rm cm^2$, which is about 300 times above current the DiDt bound. This bound weakens if $m_{\chi}>1\rm TeV$ with a heavier mediator.

The above argument suggests that one would naively expect
\beq \label{naive}
{\langle \sigma |v| \rangle_{therm} \over \sigma_{SI} } \sim 10^{5}.
\eeq
 So at least one of the assumptions we have made above must not hold in order to satisfy \eq{therm/SI}, for any realistic dark matter candidate. There are essentially two ways that \eq{therm/SI} can be satisfied, by either having $\langle \sigma |v| \rangle_{therm} = 1~{\rm pb}$ as a result of some kinematic enhancement or having some suppression for $\sigma_{SI}$.

\subsection{Basic Mechanisms Affecting ${\langle \sigma |v| \rangle_{therm} \over \sigma_{SI}}$}

For some models, a dark matter candidate does not naturally satisfy $\langle \sigma|v| \rangle_{therm}= 1~{\rm pb}$ and requires some enhancement in order to yield a thermal relic abundance. This simultaneously modifies the natural ratio in \eq{therm/SI}. The two basic mechanisms are:
\\

\underline{{\bf S-Channel Resonance}}: If dark matter annihilation is enhanced by an s-channel resonance, \eq{therm/SI} can be satisfied because of kinematic suppression in the denominator of \eq{ann}. This solution requires one to tune the mass relation between DM and mediator: $2m_{DM}\sim M$. With highly tuned $2m_{DM}=M$, resonance enhancement depends on the width-to-mass ratio, $\sim\frac{M^2}{\Gamma_M^2}$. As an example consider that for a Z-resonance, this factor is about $10^{3}$ and for a Higgs resonance ($m_{higgs}=100-200\rm GeV$), this enhancement ranges from $10^{4}-10^{10}$. In practice, some models require tuning at the few percent level to satisfy \eq{therm/SI}.

\underline{{\bf Coannihilation}}: If dark matter coannihilates with other states close to its mass,  \eq{therm/SI} can be satisfied because  $\langle \sigma |v| \rangle_{therm}$ will be dominated by coannihilation with its degenerate partner, rather than self annihilation. This also requires tuning mass relations in the model so that DM has partners with almost degenerate masses. It is most useful when DM self-annihilation is $v^2$ suppressed, while coannihilations are unsuppressed. Then this can increase $\langle \sigma |v| \rangle_{ann}$  by about $1/v^2\sim30$. This is typically not enough to solve the $O(100)$ tension we mentioned above, and therefore suppressions of $ \sigma_{SI} $ are generally required.
    %and this can be a solution for pure Bino LSP.

If a dark matter candidate does naturally have $\langle \sigma |v| \rangle_{therm} = 1~{\rm pb}$, then one or a combination of the following mechanisms usually explains how \eq{therm/SI} is satisfied. The basic principle underlying these mechanisms is the absence or suppression of the SI coupling to light quarks. Since \eq{therm/SI} arises from SI DiDt experiments, it is possible that the only operators governing annihilation, when crossed, give rise to SD scattering or do not couple to light quarks at all. While the crossed annihilation channel may not give rise to SI scattering, other operators typically do generate SI scattering. In these cases one still expects that $\langle \sigma |v| \rangle_{therm} = 1~{\rm pb}$ implies \eq{naive} if SI scattering is generated by other operators.  Of course the leading operators that do mediate SI scattering should still obey DiDt bounds. Therefore, even if the crossed annihilation diagram does not mediate SI scattering, one of the following mechanisms is typically required to suppress $\sigma_{SI}$.

\underline{ {\bf Kinematic Suppression of SI Coupling}}:
   Kinematic differences for annihilation vs. scattering can suppress $\sigma_{SI}$ relative to $\langle \sigma |v| \rangle_{therm}$. Suppressions can arise for certain operators from either the slow current-day velocity of dark matter ($\epsilon_{{\rm v}}= \left( \frac{{\rm v}_{DM}}{c} \right)^2 \sim 10^{-6}$) or the low momentum of quarks in the nucleon ($\epsilon_{QCD}=\left( \frac{\Lambda_{QCD}}{m_{DM}}\right)^2 \sim 10^{-6}$).  This can be in addition to the automatic $(\frac{m_p}{m_{\chi}})^2$ factor in \eq{mp/mdm}, which is always present. We give a brief summary and explanation of the kinematic suppressions in Appendix-A (Table-IV). More interestingly for certain operators, SD interactions have a large $\sigma_{SD}$, even when  $\sigma_{SI}$ is suppressed. Models with significant SD scattering might be probed with experiments in the near future, with exclusion limits as low as $\sigma_{SD}\sim10^{-5}\rm pb$, for instance, in ICE CUBE  \cite{Halzen:2009vu}.

\underline{{\bf Suppression from Flavor Dependent Couplings}}:  If the DM coupling to Standard Model \emph{light} quarks is suppressed so that $\sigma_{SI}$ is suppressed, then \eq{therm/SI} is readily satisfied. Meanwhile, $\langle \sigma |v| \rangle_{therm}$ remains unsuppressed in the presence of other efficient channels: heavy quarks, leptons, or W/Z. Such possibilities include:  DM coupling via a Higgs-like mediator which leads to Yukawa suppression or Z-boson mediation with mixing suppression. Yukawa suppression is a natural possibility that is incorporated in many known models. However, without incorporating additional suppression mechanisms, Yukawa suppression alone has `limited' power, and predicts $\sigma_{SI} \sim  10^{-9}-10^{-8}~{\rm pb}$ ($\epsilon_Y \sim 10^{-4}$) which will soon be probed by XENON100 and XENON1T experiments. We can understand this as follows:

Going back to eq.(\ref{ann}) and replacing $g_2$ by $y_u$ suppresses the annihilation cross-section for the u-quarks.  Taking Yukawa couplings (for all quarks) into eq.(\ref{DD0}) (replace the universal $g_2$ by $y_q$):
\beqa
     \sigma_{\chi p}  &=&\frac{1}{4\pi}\frac{m_p^2}{(m_{\chi}+m_p)^2}\frac{(g_1)^2}{M^4}
  \left(\sum_{q=u,d,s}\frac{m_p}{m_q}y_qf^p_{Tq}+\sum_{q=c,b,t}\frac{m_p}{m_q}y_q\frac{2}{27}f^p_{TG}\right)^2\label{DD2}
  \\\nonumber&\approx&\frac{1}{\pi}\frac{m_p^2}{m_\chi^2}\frac{(g_1)^2}{M^4}(\frac{m_p}{v})^2\cdot 0.2
\eeqa
with a WIMP and Higgs mass $\sim 100X~{\rm GeV}$ and $g_2=g_{EW}$, with eq.(\ref{ann}), we get  $\sigma_{DiDt}\approx \frac{m_p^2}{m_{\chi}^2}(\frac{m_p}{v})^2\frac{0.2}{g_2^2}\langle\sigma|v|\rangle_{therm}\approx 10^{-45}\rm cm^2$. So we see that this type of model is right around the reach of XENON100 or XENON1T.

\underline{{\bf Inelastic splitting}}: In this type of model, DM has a heavier `excited' partner, which opens up the possibility of inelastic scattering. If dark matter's elastic scattering to itself is suppressed or vanishes \eq{therm/SI} can be satisfied. In these cases, scattering mainly proceeds via inelastic scattering to a partner with heavier mass. Enough mass splitting between DM and its heavier partner gives a kinematic barrier, i.e. the scattering can occur only when DM carries enough kinetic energy. Therefore $\sigma_{SI}$ is suppressed by the number density distribution of DM at high velocity, which can be exponentially suppressed if SI scattering requires a very high velocity. An example of this is adding a Majorana mass which splits Dirac particle/anti-particle states into two mass eigenstates such that SI scattering is dominated by inelastic scattering. For Majorana masses larger than $1 ~{\rm MeV}$ this is sufficient to evade all DiDt bounds. With a mass splitting $\sim100\rm keV$, such models have been recently well explored for reconciling DAMA signal and null results from other DiDt experiments.

\underline{{\bf Annihilation to Dark Sector States}}:
It is possible that DM annihilates dominantly to non-Standard Model states and only couples to the Standard model via small mixing. In this case, the correlation between $\langle \sigma |v| \rangle_{therm}$ can disappear completely. Such type of models have been recently well explored in light of explaining anomalies from PAMELA, FERMI etc. \cite{ArkaniHamed:2008qn}. In these models DM annihilates dominantly to light dark sector fields which gives rise to right thermal relic density. Elastic scattering off nucleons is typically highly suppressed by small mixing between Higgs and light scalar unless proper mass splitting could accommodate a sizable inelastic scattering rate \cite{ArkaniHamed:2008qn, Finkbeiner:2008qu}

\underline{{\bf Non-Thermal Production}}:
The correlation between DiDt and annihilation cross-section disappears when dark matter is not thermal relic. We will not consider further details of these models here, but note that this simply eliminates the bound \eq{therm/SI}. Despite the popularity of thermal WIMP models, there are many well-motivated non-WIMP candidates such as the axion or gravitino LSP. Many of these models are `super-weakly' interacting both for DiDt and indirect detection. \\

\section{Some Specific WIMP Models} \label{SpecMod}

In Section \ref{ER} we have restricted to our attention to various dark matter annihilation modes as if they occur $100\%$ of the time into a particular final state with $\langle \sigma |v| \rangle_{ann} = 1 ~{\rm pb}$. However, dark matter candidates typically can annihilate into various final states in different ratios in a concrete model. In this section we select a small subset of specific dark matter candidates and consider the extent to which they can be probed by the various antideuteron searches. Each of these models avoids the naive correlation between DiDt and thermal freeze-out due to some combination of the mechanisms mentioned in the previous section. For each model, we list the leading operators that govern the particle's present day annihilation, its SI scattering, and SD scattering. We briefly describe the basic features of each model and which of the previous mechanisms is responsible for that model being kept within the bounds of DiDt experiments. We'll see that several models that are difficult to detect directly can be detected through annihilation to antideuterons.

The various cross-sections derived from dark matter/Standard Model operators may come with certain suppressions. For $\langle \sigma |v| \rangle_{ann}$, ``P-wave" ($v^2$) suppression can easily be inferred from the C and P properties of the dark matter bilinear part of the operator or CP, J conservation of initial-final states. Since the expectation value of various quark bilinears in the nucleon state determine the behavior of the SI and SD scattering cross-sections, kinematic suppressions for these can be determined simply by looking at the fermion bilinear to which dark matter couples. In what follows in this section, we will only state whether or not the leading operator has kinematic suppression and mention what type of suppression it is (Yukawa: $\epsilon_Y$, velocity: $\epsilon_{\rm v}$, or nucleon energy scale: $\epsilon_{QCD}$).  We  provide a detailed discussion of these suppression factors and their origin in Appendix A.  Also, Appendix A contains a description of the compact notation used here for listing dark matter/Standard Model interactions.

\subsection{The Models}

\begin{enumerate}

{\bf \item{ Focus-Point SUSY: Low Mass}}

\begin{center}
\begin{tabular}{|c||c|c|c|} \hline
 F.P (1) &\quad $ \bar{\chi}\gamma^{\mu}\gamma^5\chi\bar{t}\gamma_{\mu}\gamma^5t$~~(ANN) \quad &\quad $\bar{\chi}\chi\bar{f}f$~~(SI) \quad & \quad $\bar{\chi}\gamma^{\mu}\gamma^5\chi\bar{f}\gamma^{\mu}\gamma^5f$~~(SD) \quad \\
\hline
\end{tabular}
\end{center}

This SUSY dark matter candidate is mixed Higgsino and Bino having mostly Bino component. A large annihilation cross-section is achieved to top quark final states via Z-exchange (though W/Z final states are also relevant). This ``S-wave" annihilation channel is helicity suppressed for light quarks because the initial and final states must be CP odd and have $J = 0$. However, in ``S-wave," the Standard Model fermion interaction produces only the $J=1~(S=1)$ final state. A helicity flip (mass insertion) is required to conserve CP and total angular momentum. Note that for lighter DM masses the top mass does not greatly suppress the amplitude. So this cross-section determines the thermal relic abundance. SI scattering is dominated by Higgs exchange and is Yukawa suppressed. SD scattering arises from the crossed annihilation channel (with light quarks) and is not suppressed.

{\bf \item{Focus-Point SUSY: High Mass}}

\begin{center}
\begin{tabular}{|c||c|c|c|} \hline
F.P (2) &\quad $ \bar{\chi}\gamma^{\mu}\gamma^5\chi(\epsilon W\partial W)_{+\mu}$~~(ANN) \quad &\quad $\bar{\chi}\chi\bar{f}f$~~(SI) \quad & \quad $\bar{\chi}\gamma^{\mu}\gamma^5\chi\bar{f}\gamma^{\mu}\gamma^5f$~~(SD) \quad \\
\hline
\end{tabular}
\end{center}

This SUSY dark matter candidate is also mixed Higgsino and Bino but with more Higgsino component. Because the Higgsinos couple W-bosons, a large annihilation cross-section is achieved to W/Z final states (the top quark final states are also relevant). The ``S-wave"  annihilation channel is unsuppressed and determines thermal relic abundance.  SI scattering is dominated by Higgs exchange and is Yukawa suppressed. SD scattering arises from Z-boson exchange and is mildly suppressed due to a slight degeneracy between the higgsino mixture of the dark matter state \cite{Cohen:2010gj}. Essentially the SD coupling of the neutralino to the Z -boson is $\sim |Z_{H_u}|^2-|Z_{H_d}|^2$ where $Z _{H}$ is the fraction of LSP having Higgsino of a particular variety and the minus sign is from the $T^3$ generator of $SU(2)_L$. As DM gets heavier,  the Higgsino state that is the LSP is equal mixture of the two weyl Higgsinos and this couplings becomes suppressed. 

{\bf \item{ SUSY BINO:  with Coannihilation}}

\begin{center}
\begin{tabular}{|c||c|c|c|} \hline
Coann &\quad $ \bar{\chi}\gamma^{\mu}\gamma^5\chi\bar{b}\gamma^{\mu}\gamma^5b$~~(ANN) \quad &\quad $\bar{\chi}\chi\bar{f}f$~~(SI) \quad & \quad $\bar{\chi}\gamma^{\mu}\gamma^5\chi\bar{f}\gamma^{\mu}\gamma^5f$~~(SD) \quad \\
\hline
\end{tabular}
\end{center}

This SUSY dark matter candidate is Bino like. The annihilation cross-section is to $b\bar{b}$ final states, is helicity or ``P-wave"  suppressed, and too small to yield the correct relic abundance.  A thermal relic abundance is achieved via coannihilations with a mass degenerate species (usually stau) in the early universe.  Higgs exchange dominates the SI scattering and is $\epsilon_Y$ suppressed. SD scattering arises from Z-boson exchange.

{\bf \item{ SUSY BINO: with s-channel Resonance (``A-funnel")}}

\begin{center}
\begin{tabular}{|c||c|c|c|} \hline
A-funnel &\quad $ \bar{\chi}\gamma^5\chi\bar{f}\gamma^5f$~~(ANN) \quad &\quad $\bar{\chi}\chi\bar{f}f$~~(SI) \quad & \quad $\bar{\chi}\gamma^{\mu}\gamma^5\chi\bar{f}\gamma^{\mu}\gamma^5f$~~(SD) \quad \\
\hline
\end{tabular}
\end{center}

This SUSY dark matter candidate is Bino like. A large ``S-wave"  annihilation cross-section is achieved from resonant s-channed CP-odd Higgs boson exchange assuming that $m_{DM} \sim 2m_{A}$. This can yield the correct relic abundance.  The crossed annihilation diagram mediates subdominant $\epsilon_{\rm v}\epsilon_Y$ suppressed SD scattering. Dominant SD scattering for DiDt is through Z-exchange. SI scattering is via Higgs exchange, and is $\epsilon_Y$ suppressed.

{\bf \item{ UED Model: KK $B^{(1)}$}}

\begin{center}
\begin{tabular}{|c||c|c|c|} \hline
UED &\quad $ (\epsilon V \partial V)_+^{\mu}\bar{f}\gamma_{\mu}f$~~(ANN) \quad &\quad $ (\epsilon V \partial V)_+^{\mu}\bar{f}\gamma_{\mu}f$~~(SI) \quad & \quad $(\epsilon V \partial V)_+^{\mu}\bar{f}\gamma_{\mu}\gamma^5f$~~(SD) \quad \\
\hline
\end{tabular}
\end{center}

This UED  \cite{Servant:2002aq} dark matter annihilates to fermions via t-channel KK fermion exchange. ``S-wave" annihilation to quarks is unsuppressed and can yield the correct thermal relic abundance. Crossed versions of the annihilation diagrams yield SI and SD scattering \cite{Servant:2002hb}. SI scattering via this operator\footnote{ We have listed only one of the operators  one gets from decomposing the t-channel fermion exchange diagram into operators with dark matter and Standard Model billinear operators. However, the other operators have equal or more kinematic suppression. For example $(VV)_+^{\mu\nu}(\bar{f}\gamma_{\mu}\partial_\nu f)_-$ which is associated with $\epsilon_{QCD}$, while $\epsilon_{QCD}\sim\epsilon_v$ for a weak scale DM}  is kinematically suppressed because only the time component of the dark matter momentum is unsuppressed, implying that only the $\bar{f}\gamma^if$ component of the fermion part of the operator is relevant and yields $\epsilon_{\rm v}$ suppressed SI scattering. The SD scattering is not kinematically suppressed, however, since the mediator is a KK (1~{\rm TeV}) state rather that a $Z$-boson. The SD scattering is roughly $10^{-2}$ of the typical SUSY SD scattering.

{\bf \item{ Little Higgs with T-Parity Model}}

\begin{center}
\begin{tabular}{|c||c|c|c|} \hline
LHTP &\quad $ (VV)(WW)$~~(ANN) \quad &\quad $VV\bar{f}{f}$~~(SI) \quad & \quad $(\epsilon V \partial V)_+^{\mu}\bar{f}\gamma_{\mu}\gamma^5f$~~(SD) \quad \\
\hline
\end{tabular}
\end{center}

Little Higgs models with T-Parity (LHTP)  \cite{Cheng:2004yc,Cheng:2003ju,Hubisz:2004ft} contains a spontaneously broken ``mirror" gauge symmetry and the dark matter is a massive ``mirror" photon of its $U(1)$ factor. In these models dark matter can  ``S-wave" annihilate through a Higgs resonance and produce the correct thermal relic abundance, but the natural annihilation cross section is typically too small and requires $m_{DM} = 2m_{Higgs}$ in order enhance the s-channel annihilation to the correct value. This induces an s-channel resonance suppression of the ratio: $\frac{\langle \sigma |v| \rangle_{therm}}{\sigma_{SI}}$. Since the annihilation is not to fermions the SI and SD scattering are determined by different operators than annihilation \cite{Birkedal:2006fz}. The SI  interaction is dominated by Higgs exchange and therefore gets a Yukawa suppression in addition to the suppression relative to the s-channel resonance . The SD interaction is similar as in UED in addition to another factor of $10^{-4}$ suppression. This suppression arises from the fact that in Little Higgs Models with T-parity, the Standard Model is charged under a spontaneously broken gauge group with ``mirror hypercharge" ($Y'=\frac{1}{10}$); since the dark matter is the massive photon of that spontaneously broken symmetry, then it's SD elastic scattering cross section off quarks is suppressed  by $Y'^4$.

{\bf \item{ Warped Extra-dimensional Model: KK LZP $\nu_R^{(1)}$}}

\begin{center}
\begin{tabular}{|c||c|c|c|} \hline
LZP &\quad $\bar{\chi}\gamma^{\mu}\chi\bar{t}\gamma_{\mu}t$~~(ANN) \quad &\quad $\bar{\chi}\gamma^{\mu}\chi\bar{f}\gamma_{\mu}f$~~(SI) & \quad $\bar{\chi}\gamma^{\mu}\gamma^5\chi\bar{f}\gamma_{\mu}\gamma^5f$~~(SD) \quad \\
\hline
\end{tabular}
\end{center}

In models of warped extra dimensions it is possible to have dark matter that couples mainly to tops \cite{Agashe:2004ci,Agashe:2004bm} or more generally, a Dirac fermion with flavour dependent couplings to quarks via Z' gauge bosons \cite{Jackson:2009kg}. In these cases, annihilation to $\bar{t}t$ is unsuppressed and gives the correct relic abundance. SI scattering is not kinematically suppressed, but is small due to a small Z/Z' mixing. SD scattering is suppressed in the same way. This suppression is similar in principal to ``Yukawa" suppression where flavor dependent couplings play an important role.

{\bf \item{ Neutral Singlet Scalar Mixing with Higgs}}

\begin{center}
\begin{tabular}{|c||c|c|c|} \hline
Scalar &\quad $ \phi^{\dagger} \phi hh$~~(ANN) \quad &\quad $\phi^{\dagger}\phi \bar{f}f$~~(SI) & \quad $0$~~(SD) \quad \\
\hline
\end{tabular}
\end{center}

One of the simplest dark matter models of all is a neutral stable scalar having quartic interactions with the Higgs boson \cite{McDonald:1993ex,Burgess:2000yq,Ponton:2008zv}.
%In this model annihilation can be to Higgs bosons,
For mass above $m_h$, DM annihilates to Higgs bosons, resulting in the correct thermal relic abundance. W-bosons can be an important final state for some scenarios.
The SI scattering is Yukawa suppressed and there is no SD scattering.

{\bf \item{ Doublet/Singlet Scalar Inelastic Model}}

\begin{center}
\begin{tabular}{|c||c|c|c|} \hline
Doublet/Scalar &\quad $ (\phi^{\dagger}\partial\phi)_-^{\mu}\bar{f}\gamma_{\mu}f$~~(ANN) \quad &\quad $0$~~(SI) & \quad $(\phi^{\dagger}\partial\phi)_-^{\mu}\bar{f}\gamma_{\mu}\gamma^5f$~~(SD) \quad \\
\hline
\end{tabular}
\end{center}

A light scalar doublet/singlet mixture whose mass is just below the W-mass threshold can provide a correct relic abundance. This model is a variant of models appearing in \cite{Cui:2009xq}. We will leave a discussion of the details of this model for Appendix B and only mention the essential features here. The dominant annihilation in the early universe is to quarks and this is ``P-wave" suppressed, but for light enough dark matter can yield the correct relic abundance. Present day annihilation is dominated by annihilation to two (slightly) off-shell W-bosons so that $\langle \sigma |v| \rangle_{ann} = 0.1 ~{\rm pb}$.  SI elastic scattering is absent, but inelastic scattering is possible. This feature can be used to explain the apparent conflict between DAMA annual modulation signal and null discovery in other DiDt experiments\cite{TuckerSmith:2001hy}. General strong neutrino bounds on inelastic DM from solar capture (\cite{Menon:2009qj}) are somewhat relieved in this model as a result of lighter mass and mixing with singlet. In this particular model the SD scattering is mediated by Z-boson exchange and is of typical size.

{\bf \item{SUSY Wino: non-thermal}}
\end{enumerate}

\begin{center}
\begin{tabular}{|c||c|c|c|} \hline
Wino  &\quad $ \bar{\chi}\gamma^{\mu}\gamma^5\chi(\epsilon W\partial W)_{+\mu}$~~(ANN) \quad &\quad $\bar{\chi}\chi\bar{f}f$~~(SI) \quad & \quad $\bar{\chi}\gamma^{\mu}\gamma^5\chi\bar{f}\gamma^{\mu}\gamma^5f$~~(SD) \quad \\
\hline
\end{tabular}
\end{center}
Finally, we consider one dark matter candidate that does not have present day cross section equal to that of a thermal relic, a Wino-like SUSY LSP \cite{Moroi:1999zb}.   It has a large present day annihilation cross-section. SI scattering arises from Higgs exchange and is Yukawa suppressed \cite{Nagai:2008se}. SD coupling arises from Z-boson exchange and is also suppressed \cite{Hisano:2010fy}. This model predicts a very large antideuteron signal for all experiments mainly due to its large annihilation cross-sections ($10^{2} ~{\rm pb}$). It is plausible that propagation parameters other than those quoted in this paper can be used for the astrophysical background \cite{Grajek:2008pg,Kane:2009if} and can be in agreement with anti-proton backgrounds. While relaxing the allowed propagation of background antiprotons one does not significantly reduce the propagation of dark matter antiprotons or antideuterons. Thus a Wino candidate, such as this one, should be robustly detected or ruled out by future antideuteron searches.

\subsection{Model Summary}

 The leading present day annihilation channels for all of these models are not ``P-wave" suppressed and have significant annihilation to final states that produce hadrons. Therefore they will contribute to the population of antideuteron cosmic rays. Picking particular parameters for each model, we can use our previous general analysis to determine each model's dark matter contribution to the antideuteron cosmic ray spectrum and calculate an expected number of antideuterons for the GAPS and AMS-02 experiment. Our results are summarized in Table \ref{specmod}. We list the mass and present day annihilation cross-section along with the fraction of annihilations to final states yielding hadrons. For the supersymmetric models we have used the numerical package DARKSUSY 5.0.4 \cite{Gondolo:2004sc,darksusy2} to check that the choice of parameters yields the correct relic abundance. The choices of parameters are similar to the parameter space described in \cite{ArkaniHamed:2006mb} for the Focus Point regions of parameter space and \cite{Baltz:2006fm} for the Coannihilation and ``A-funnel"  regions of parameter space. We list the predicted number of antideuteron events for the GAPS (ULDB) experiment as well as an estimate for a satellite mission, GAPS (SAT). The number of events predicted should be compared to the $N_{crit}$ column of Table \ref{reach} in order to determine whether or not the predicted signal is significant. Finally, we list the expected order of magnitude for SI/SD elastic scattering off of nucleons, which reflects the kinematic and coupling suppressions mentioned above.

 We have not included the AMS-02 and GAPS(LDB) experiments in this list because they are not sensitive enough to detect most of the models for the parameters chosen here. However,  Wino dark matter is a notable exception. For $m_{DM} = 200 ~{\rm GeV}$, this predicts 10 (12) events in the high (low) energy AMS-02 experiment and 8 events in the GAPS (LDB) experiment. Similarly, for $m_{DM} =300 ~{\rm GeV}$, this predicts 2 (1) events in the high (low) energy AMS-02 experiment, and 1 event in the GAPS (LDB) experiment. While AMS-02 and GAPS(LDB) are not promising for probing the particular thermal relics discussed in Table \ref{specmod}, based on our model independent analysis in Section \ref{ER}, we can see that the mass reach for AMS-02 is $150~{\rm GeV} (110~{\rm GeV})$ for the low (high) energy window. Also, while GAPS(ULDB) will have greater sensitivity than AMS-02, AMS-02 will remain the only probe of antideuterons with kinetic energies $ T_{\bar{D}} \geq 1~{\rm GeV}$. 

\begin{table}[t!]
\centering
\begin{tabular}{|c||c|c|c|c|c|c|c|c|c|c|} \hline
Model                       & $m_{DM}$     &$\sigma|v|/{\rm pb}$& $\xi_W$ & $\xi_{q}$ & $\xi_{t}$&$\xi_{h}$& $N_{3\sigma}=2$ & $N_{3\sigma}=10$& $\sigma_{SI}/n/{\rm pb}$&$\sigma_{SD}/n/{\rm pb}$\\
                                  & $({\rm GeV})$&                                &                 &                 &               &                           &  $N_{5\sigma}=5$        &  $N_{5\sigma}=16$       &        &  \\
                                   &                          &                                &                 &                 &               &                           &   (ULDB)         &  (SAT)        &       &  \\ \hline

SUSY F.P (1)         &190                   & $0.67$                & $0.2$          & $0.02$    & $0.73$   & $0$                       & 4                 & 50                     & $10^{-8} $                & $ 10^{-4}$\\ \hline
SUSY F.P  (2)       &772                   & $0.33$                 & $0.55$       &  $0$         &  $0.38$   &$0$                        & 0                  & 4                     & $10^{-8} $                & $ 10^{-5}$\\ \hline
SUSY coann        &148                 & $0.17$                    & $0$          & $1$     &  $0$       &$0$                        & 1                  & 14                     & $10^{-8} $                & $10^{-6}$\\ \hline
SUSY A-funnel   &163                 & $0.6$                       & $0$          & $0.92$     &  $0$        &$0$                         & 2                 & 33                      & $10^{-8} $                  &$  10^{-6}$\\ \hline
UED $B^{(1)}$          &900                & $0.6$			& $0$  	& $0.19$ 	     &  $0.16$    &$0.02$    		 & $0$ 	     & $3$ 		     & $10^{-8}$		  &$10^{-6}$ \\ \hline
UED $B^{(1)}$ coann.& 600	      &$0.6$		& $0$  	& $0.19$       &$0.16$      & $0.02$   		& $0$   	     &  $4$  		     &  $10^{-8}$  		 &$  10^{-6}$ \\ \hline
%UED $\nu^{(1)}$    &1300		     &$0.17$	& $0.46$  & $0$ 	 & $0$     & $0$  	          & $0$   	      &  $0$  	     &  $0$		  & $\sim 10^{-3}$  \\ \hline
%UED $\nu^{(1)}$ 3-gen. &950	    &$0.17$	& $0.46$  & $0$ 	&  $0$     &$0$   	          & $0$   	     &  $0$  		     & $ 1$    	 &$ \sim 10^{-3}$\\  \hline
LHTP  		      &200		    & $0.8$ 		& $1$ 	&  $0$ 	    &  $0$      &$0$   		& $0$   	     &  $12$  		     & $10^{-12}-10^{-10}$ 	 & $ 10^{-10}$\\  \hline
LZP $\nu^0_R$ &300	   	     & $1$       		& $0.06$   & $0$           & $0.94$      &  $0$   		& $3$  	     &  $41$ 		     & $ 10^{-9}-10^{-6}$ 	&$ 10^{-7}-10^{-4}$ \\ \hline
Singlet (scalar)  &200		     & $1$ 			& $0$ 	& $0$         &  $0$     & $1$   		& $2$   	     &  $36$  		     & $ 10^{-8}$ 	 &$0$ \\  \hline
Doublet/Singlet & 75		     & $0.1$ 			& $1$ 	& $0$ 	  &   $0$    &$0$   		& $3$   	      &  $49$  	     & $ 0$ 		& $ 10^{-4}$ \\ \hline
Wino (non-therm)  &200           			     & $65$ 		& $1$ 	&  $0$ 	 &  $0$      &$0$          	& $50$   	      &  $730$   	     & $10^{-12}-10^{-8}$     & $<10^{-11}$\\  \hline
Wino (non-therm)  &300           			     & $30$ 		& $1$ 	&  $0$ 	 &  $0$      &$0$          	& $5$   	      &  $75$   	     & $10^{-12}-10^{-8}$     & $<10^{-11}$\\  \hline
\end{tabular}
\caption{\label{specmod} Predicted number of antideuterons detected in various experiments for different dark matter models. Here $\xi_f $, where ``$f$" is some final state, signifies the fraction dark matter annihilations that yield this particular final state. Also, $\xi_q$ stands for all light quarks (including b-quarks) and $\xi_t$ includes only top-quarks. Otherwise these are the same parameters as in Table \ref{reach}. We also list the typical range for SI and SD elastic scattering off nuclei.}
\end{table}

\section{Conclusions} \label{conc}

In this paper, we have considered the mass reach of the antideuteron cosmic ray search experiments AMS-02 and GAPS. The basic point can be summarized by stating that for typical thermal dark matter annihilating with a present day annihilation cross-section of  $\langle \sigma |v| \rangle_{ann} = 1~{\rm pb}$, GAPS (ULDB) expects to detect antideuterons of dark matter origin at greater than $99.7\%$ confidence for masses up to $m_{DM} = 250 ~{\rm GeV} $ in the case of annihilation to $h^0h^0$, $m_{DM} = 400 ~{\rm GeV} $ for $\bar{t}t$, $m_{DM} = 250 ~{\rm GeV} $ for $\bar{q}q$, $m_{DM} = 160 ~{\rm GeV} $ for $\bar{W}W$, and $m_{DM} = 500 ~{\rm GeV} $ for $\bar{g}g$ final states, assuming MED propagation parameters. At $5 \sigma (\sim 99.99\% C.L.)$  GAPS (ULDB) expects to discover antideuterons of dark matter origin for DM masses up to $m_{DM} = 150 ~{\rm GeV} $ for $h^0h^0$, $m_{DM} = 220 ~{\rm GeV} $ for $\bar{t}t$, $m_{DM} = 150 ~{\rm GeV} $ for $\bar{q}q$, $m_{DM} = 120 ~{\rm GeV} $ for $\bar{W}W$, and $m_{DM} = 300 ~{\rm GeV} $ for $\bar{g}g$ final states. Therefore antideuterons are a significantly more sensitive way to probe the hadronic annihilation channels of dark matter than antiprotons. A GAPS experiment on a satellite, obtaining an order of magnitude improvement in sensitivity, would provide the same mass reach at $5\sigma$ significance that GAPS (ULDB) achieves for $2\sigma$ significance and could see primary antideuterons for dark matter masses of $1~{\rm TeV}$.  GAPS (ULDB) is virtually background free, and satellite level sensitivities are expected to see only a few background events. AMS-02 and the GAPS (LDB) are also sensitive to antideuteron cosmic rays of dark matter, and the mass reach is roughly $m_{DM} = ~100{\rm GeV}$.

We have computed injection spectra according to the non-factorized coalescence model \cite{Kadastik:2009ts} which more accurately takes into account angular correlations in antideuteron formation when hadronization is modeled with PYTHIA. Since this has, correctly, increased the prediction in antideuterons at higher energies we have produced flux results that are somewhat flatter at higher energies than previous analysis. While this does not make a great difference for low energy antideuteron searches, it does affect the higher energy AMS-02 detection rates. For a fixed annihilation rate of $\langle \sigma |v| \rangle_{ann} = 1 ~{\rm pb}$ and MED parameters, we find a $100 ~{\rm GeV}$ dark matter candidate predicts 1 event for annihilation to $\bar{b}b$. Since dark matter can be as light as $50~{\rm GeV}$ without violating the antiproton bound, this high energy band may still be relevant for dark matter searches.

The dominant uncertainty comes from uncertainties associated with cosmic ray propagation from dark matter annihilation . The minimum flux can be an order of magnitude less than the flux inferred from using the MED parameters used to make the plots in this paper. This results in degrading the mass reach by a factor of 3. However, the mass reach, while less, remains an order of magnitude better than the bound from antiprotons. Because the model of antideuteron production is fit to the ALEPH measurement of Z-boson decay to antideuterons, we believe the hadronization uncertainties are $\mathcal{O}(1)$, which are small compared to propagation errors.

We have also described the general tension between dark matter as a thermal relic and bounds set by Direct Detection experiments. We have listed the basic mechanisms that are at work in many different models of dark matter responsible for eliminating this superficial conflict. These mechanisms typically manifest themselves as different suppression factors for SI scattering. In Appendix A we tabulated these suppression factors associated with different Standard Model fermion bilinears, when their expectation value is taken in a nucleon state.  We then used Appendix A to understand different dark matter models by their leading dark matter/Standard Model operators which determine annihilation and elastic scattering off nucleons. We identified which mechanisms/suppression factors are present in each model alleviating the conflict between the dark matter as a thermal relic and  Direct Detection bounds. Finally, for each of these models, we have picked a particular dark matter mass (and model parameters) that satisfies $\Omega h^2 =0.12$ and determined the antideuteron signal predicted in the GAPS (ULDB) and GAPS (SAT) experiment. Our selection of models is meant to be representative of typical models but by no means complete.  We expect that GAPS (ULDB) will provide a first real test for hadronic annihilation channels for dark matter with masses smaller than $100 ~{\rm GeV}$. For GAPS (SAT), the possibility exists to probe hadronic annihilation for masses up to $m_{DM} = 1~{\rm TeV}$.

\section*{Acknowledgements}

We have benefited from conversations with Doug Finkbeiner, Tongyan Lin, Rene Ong, David Poland, Matthew Schwartz, Neil Weiner, Itay Yavin. Lisa Randall would like to thank NYU for their hospitality while this work was being completed. This work was supported by NSF grant PHY-0855591 and also by the Harvard Center for the Fundamental Laws of Nature.

\appendix

\section{Operator Properties Relevant for Dark Matter Detection}

As commented in section IV, a general DM candidate's annihilation cross-section, SI scattering, and SD scattering depend on the operators involved in interactions with the Standard Model (e.g. whether there is $v^2$ suppression in SI DiDt rate or IdDt rate). In this Appendix we shall classify dark matter scalar, fermionic (Dirac and Majorana), and vector boson bilinears according to their transformation properties under Charge Conjugation (C)  and Parity (P) and derive kinematic suppressions for the expectation value of bilinear operators between the vacuum and a non-relativistic ($\frac{v}{c} \ll 1$) particle anti-particle state. This will be relevant for determining whether ``S-wave," ``P-wave," or helicity suppression plays a role in dark matter annihilation. We present this for completeness; much of what appears here can be found in standard texts \cite{Weinberg:1995mt,Peskin:1995ev,Srednicki:2007qs}. We will also tabulate kinematic suppression factors for quark bilinears in a nucleon state, which will be important for interpreting bounds from Direct Detection (DiDt) experiments.

We list a basis of bilinear operators for fermions in Table \ref{ferm}, for scalars in Table \ref{sca}, and for gauge bosons in Table \ref{vec}. These tables are to be used in Section \ref{SpecMod} to infer the kinematic suppressions for annihilation or SI/SD scattering for particular dark matter/Standard Model operators. These tabulated bilinear operators can be interpreted as being dark matter or Standard Model field bilinears. All possible 4 point interactions between Standard Model and dark matter can be written as a product of two of the bilinear operators tabulated here in form of $\mathcal{O}_{DM}\mathcal{O}_{SM}$. Any dark matter/Standard Model operator can be brought into this form. For example,  $(\bar{\chi}(a+b\gamma_5)f)(\bar{f}(a-b\gamma_5)\chi)$ for SUSY neutralino-SM fermion interaction via t-channel exchanging sfermion, can be reorganized using Fierz identities to a combination of operators of the form described above. Restricting to CP-conserving interactions simplifies the number of 4 point interactions, however the number of operators is still quite large. We restrict our attention to interaction operators up to dimension-6. Some short-hand notations are used in order to fit into a compact table, in particular for vector boson fields, which will be explained. One common notation in form of $(\bar{\chi}A\chi)_\pm$ is short-hand for two bilinears with definite C parity: $\bar{\chi}A\chi+\chi A\bar{\chi}$, $i(\bar{\chi}A\chi-\chi A\bar{\chi})$, respectively. Obviously for a real field only the $+$ one survives. Another $(-)^{\mu,\nu}\equiv(-1)^{\mu}(-1)^{\nu}$ where $(-1)^0=+1, (-1)^i=-1 (i=1,2,3)$.\\

In the case that bilinears are of fields describing dark matter, it is a simple exercise to establish whether or not present day annihilation is ``P-wave" suppressed or not based on C and P properties of the bilinear operator and  and ``S-wave" state (at the end of this appendix we comment on some particular cases where CP, J conservation between initial and final state needs to be considered to determine ``P-wave" suppression or helicity suppression at ``S-wave").

Consider the massive particle/anti-particle state: $| p_1,l_1, s_1;p_2,l_1, s_2 \rangle$.  This state can be decomposed into a superposition of orbital angular momentum eigenstates. Since dark matter annihilates in the non-relativistic limit, we will neglect all but the $l = 0,1$ states, also called ``S-wave" and ``P-wave,"  when the states are initial states of a $2 \rightarrow 2$ annihilation process. For ``S-wave" or ``P-wave" states, the state then has well defined $C$ and $P$ transformation properties given by:

\beqa
&&C=(-1)^{L+S}, P=(-1)^L   ~ ({\rm Scalars/Gauge Bosons}) \\
&&~~~C=(-1)^{L+S}, P=(-1)^{L+1}~({\rm Fermions})  ~~\\
\eeqa
Where ``L" is the orbital angular momentum quantum number and ``S" is the total spin quantum number. Since dark matter annihilation to the standard model will always involve expectation values of the form:
\beq \label{op}
\langle  p_1,l_1, s_1;p_2,l_1, s_2 | \mathcal{O}(\bar{\chi},\chi) |0\rangle
\eeq
where $\mathcal{O}$ is some dark matter bilinear operator. Then we may immediately use the $C$ and $P$ properties of the state and the operator to determine whether these expectation values are zero or not for an ``S-wave" initial state; in order to be non-zero the quantum numbers of the state and operator should match.

%One simply needs to check that the C and P of a dark matter initial state may be the same as the bilinear operator.
For example, the fermion bilinear $\bar{\chi}\chi$ always contributes ``P-wave" suppressed annihilation because there does not exist a fermion/anti-fermion state that is C and CP even with zero orbital angular momentum. In the tables  `` $\checkmark$" indicates that \eq{op} is non-zero for an ``S-wave"  state and a given dark matter billinear. Furthermore, it is not uncommon for dark matter candidates to be their own anti-particles. In this case all two particle dark matter states are even under Charge Conjugation and only the operators with ``+" subscripts yield non-vanishing contributions to either annihilation or SI/SD scattering.

If one considers the case that the fermions in the fermion billinears of Table \ref{ferm} are Standard Model light quarks, then the expectation value of these operators in a nucleus state determines important features for $\sigma_{SI/SD}$: whether it is SD or SI and whether it has certain type of suppression as indicated by $\epsilon_Y, \epsilon_{\rm v}, \epsilon_{QCD}$ as explained below ($\epsilon_Y, \epsilon_{\rm v}$ also apply when the fermion in the bilinear is DM that scatters off nucleus.). In the first two rows (``SI', ``SD") of Table \ref{ferm} we summarize the contribution of the operator for momentum transfer ($q$) which for DiDt is typically $q \propto v \sim O(\rm keV)\ll\rm m_N, m_{DM}$.  $\checkmark$ means no suppression; $0$ indicates strictly vanishing; $\epsilon_{\rm v}$ indicates a suppression $\sim v^2\sim10^{-6}$ for $\sigma_{DiDt}$; $\epsilon_{QCD}$ indicates a suppression $\sim(\frac{\Lambda_{QCD}}{m_{DM}})^2( \sim10^{-6}~ {\rm for}~ m_{DM}\sim100\rm GeV)$; $\epsilon_Y$ indicates that a possible Yukawa suppression $\sim (\frac{m_N}{v_{EW}})^2\sim10^{-4}$ if the operator originates from integrating out a Higgs-like mediator. It can be seen that typically either SI or SD has $\epsilon_{\rm v}$ suppression, but not both, which results from the fact that in the $q\rightarrow0$ limit either $\sigma^0$ or $\sigma^i$ part of matrix element is picked up. Details can be found in \cite{Kurylov:2003ra, Agrawal:2010fh}.

\begin{table}
\begin{center}
   \begin{tabular}{|c|c|c|c|c|c|c|c|c|}
     \hline
      &{\small{$\bar{\Psi}\Psi$}} & {\small{$\bar{\Psi}\gamma^5\Psi$}} & {\small{$\bar{\Psi}\gamma^\mu\Psi$}} & {\small{$\bar{\Psi}\gamma^\mu\gamma^5\Psi$}} & {\small{$\bar{\Psi}\sigma^{\mu\nu}\Psi$}} & {\small{$\bar{\Psi}\sigma^{\mu\nu}\gamma_5\Psi$}} & {\small{$(\bar{\Psi}\gamma^\mu\partial^\nu\Psi)_\pm$}} & {\small{$(\bar{\Psi}\gamma^\mu\gamma^5\partial^\nu\Psi)_\pm$}} \\
    \hline SI & $\epsilon_Y$ & $0$ & $\checkmark$ & $\epsilon_{\rm v}$ & $\epsilon_{\rm v}$ & $\epsilon_{\rm v}$& $\epsilon_{QCD}$ & $\epsilon_{\rm v}$ \\
    \hline SD & $0$ & $\epsilon_{\rm v} \epsilon_Y$ & $\epsilon_{\rm v}$ &$\checkmark$ & $\checkmark$ & $\checkmark$ & $\epsilon_{\rm v}$ & $\epsilon_{QCD}$\\
     \hline C & $+$ & $+$ & $-$ & $+$ & $-$ & $-$ & $\mp$ & $\pm$ \\
     \hline P & $+$ & $-$ & $(-)^\mu$ & $-(-)^\mu$ & $(-)^{\mu,\nu}$ & $-(-)^{\mu,\nu}$& $(-)^{\mu,\nu}$ & $-(-)^{\mu,\nu}$ \\
   %  \hline ? H & LL, RR & LL, RR & LR & LR & LL, RR &  & LR & LR \\
     \hline s-wave & 0 & $\checkmark$ & $\checkmark$ & $\checkmark$ & $\checkmark$ & $\checkmark$ & $+:\checkmark, -: 0$  & $+: 0, -: \checkmark$ \\
   %  \hline p-wave & $\checkmark$ & 0 & $\checkmark$ & $\checkmark$ & $\checkmark$ & & $\checkmark$ & $\checkmark$ \\
     \hline
   \end{tabular}
   \caption{\label{ferm}Here is a list of fermion billinears, up to dimension four, having definite C and P transformation properties. Listed is each operator's suppression factor when its expectation value is taken in a nucleon state. Also listed is whether or not  the operator allows``S-wave" annihilation for a dark matter state. }
   \end{center}
   \end{table}

\begin{table}
\begin{center}
\begin{tabular}{|c|c|c|c|}
  \hline
 & $\phi^\dag\phi$ & $(\phi^\dag\partial_\mu\phi)_\pm$& $(\phi^\dag\partial_\mu\partial_\nu\phi)_\pm$\\
 \hline C & $+$ & $\pm$ & $\pm$\\
  \hline P & $+$ & $(-)^\mu$ &$(-)^{\mu,\nu}$ \\
%\hline  J & $0$ & $1$ & & &\\
\hline s-wave& $\checkmark$ & $+: \checkmark, -: 0$ & $+: \checkmark, -: 0$\\
%  \hline p-wave& $\checkmark$ & $\checkmark$ & \\
  \hline
\end{tabular}
  \caption{\label{sca}Here is a list of scalar billinears, up to dimension four, having definite C and P transformation properties. Listed is whether or not  the operator allows``S-wave" annihilation for a dark matter state.  }
   \end{center}
\end{table}

\begin{table}
\begin{center}
\begin{tabular}{|c|c|c|c|c|c|c|c|c|c|}
  \hline
   & {\small{$V V$}} & \small{{$(VV)^{\mu\nu}_{\pm}$}} & {\small{$(\epsilon V V)^{\mu\nu}_\pm$}} & {\small{$(V\partial V)^\mu_\pm$}} & {\small{$(\epsilon V\partial V)^\mu_\pm$}} & {\small{$(V\partial\partial V)^{\mu\nu}_\pm$}} & {\small{$(\epsilon V\partial\partial V)^{\mu\nu}_\pm$}}& {\small{$(V\partial^2 V)_\pm$}}& {\small{$(\epsilon V\partial\partial V)_\pm$}}\\
  \hline {\small{ C}} & $+$ & {$\pm$}& $\pm$& $\pm$&$\pm$ & $\pm$& $\pm$& $\pm$&$\pm$\\
  \hline {\small{P}} & $+$ & {\small{$(-)^{\mu,\nu}$}} & {\small{$-(-)^{\mu,\nu}$}} & {\small{$(-)^\mu$}} & {\small{$-(-)^\mu$}}& {\small{$(-)^{\mu,\nu}$}}& {\small{$-(-)^{\mu,\nu}$}}& {\small{$+$}}& {\small{$-$}}\\
 %\hline H & L$\uparrow$ L$\downarrow$, T$\uparrow$ T$\downarrow$ & LT & LT & L$\uparrow$ L$\downarrow$, T$\uparrow$ T$\downarrow$ & & & & & \\
  \hline {\small{s-wave}} & $\checkmark$ & $\checkmark$ & $\checkmark$ & \multicolumn{6}{|c|}{$+: \checkmark, -:0$}\\
%  \hline p-wave & 0 & $\checkmark$ & $\checkmark$ & $\checkmark$ & & & & &\\
  \hline
\end{tabular}
\caption{\label{vec}Here is a list of Massive Spin-1 billinears, up to dimension four, having definite C and P transformation properties. Listed is whether or not  the operator allows ``S-wave" annihilation for a dark matter state.  }
\end{center}
\end{table}

For vector boson billinears we adopt the following notation:
\beqa
&&VV\equiv V^+_\mu V^-_\mu,~~~~
 (VV)^{\mu\nu}\equiv V^{+\mu}V^{-\nu}, ~~~~
(\epsilon V V)^{\mu\nu}\equiv\epsilon^{\mu\nu\rho\sigma} V^+_\rho V^-_\sigma,\\\nonumber
&&(V\partial V)^\mu\equiv V^{+\mu}\partial^\nu V_{-\nu}, V^{+\nu}\partial^\mu V_{-\nu}, V^{+\nu}\partial_\nu V^{-\mu}, ~~~~(\epsilon V\partial V)^\mu\equiv\epsilon^{\mu\nu\rho\sigma}V_{+\nu}\partial_\rho V_{-\sigma}, \\\nonumber
 &&  (V\partial\partial V)^{\mu\nu}\equiv V^{+\mu}\partial^\nu\partial^\rho V^-_\rho, V^{+\mu}\partial^2 V^{-\nu}, V^{+\mu}\partial_\rho\partial^\rho V^{-\nu}, V^{+\rho}\partial_\rho\partial^\mu V^{-\nu},\\\nonumber
 && (\epsilon V\partial\partial V)^{\mu\nu}\equiv\epsilon^{\mu\nu\rho\sigma}V^{+}_{\rho}\partial_\sigma\partial_\lambda V^{-\lambda}, \epsilon^{\mu\nu\rho\sigma}V^{+}_{\rho}\partial^2 V^{-\sigma}, \epsilon^{\mu\nu\rho\sigma}V^{+}_{\lambda}\partial_\rho\partial_\sigma V^{-\lambda}\\\nonumber
 &&V\partial^2 V\equiv V^{+\mu}\partial^2V^{-}_{\mu}, ~~~~\epsilon V\partial\partial V\equiv\epsilon^{\mu\nu\rho\sigma}V^{+}_{\mu}\partial_\nu\partial_\rho V^-_\sigma
 \eeqa

A dark matter/Standard Model operator of the form $\mathcal{O}_{DM}\mathcal{O}_{SM}$ is in general a combination of two of the bilinears listed above. Even if one restricts to CP conserving operators, the list of possible operators up to dimension six is well above 50. Therefore we do not list these operators here. However, given such an operator, kinematic suppressions for DiDt can be read off directly by looking at which SM fermion bilinear appears in the operator and reading off its suppression factor from Table \ref{ferm}. Similarly, in order to determine whether or not annihilations are ``P-wave" suppressed, one typically only needs to determine if the cooresponding dark matter bilinear has has non-zero expectation value in a two particle ``S-wave" dark matter state.  However, in some cases this information in combination with selection rules from CP and J conservation is required in order determine whether or not helicity suppression is present. We give two brief examples here. First, in order to understand ``S-wave" annihilation via the operator $\bar{\chi}\gamma^\mu\gamma^5\chi\bar{f}\gamma^\mu(\gamma^5)f$ (an important operator for SUSY Bino LSP annihilation), notice that $\bar{\chi}\gamma^\mu\gamma^5\chi$ has $C=+1$ and fermion pair state has $P=-1$ and determines the initial state should be at $CP=-1, S=0$ state at $L=0$. To conserve $J=0$ final state needs to have opposite spin orientation. But for massless fermion amplitude for this helicity structure vanishes for state connected to $\bar{f}\gamma^\mu(\gamma^5)f$. Therefore a mass insertion is needed to flip helicity, which implies a Yukawa-like suppression for annihilation to light SM fermions. Another example is, $\bar{\chi}\gamma_\mu\gamma^5\chi(V\partial V)^\mu_-$. Again for the same reason as in the first example, at $L=0$ initial fermion state should be at $J=0, C=+1, P=-1$, which selects $\mu=0$ component of $\bar{\chi}\gamma^\mu\gamma^5\chi$ to satisfy $P=-1$. Accordingly $\mu=0$ or ``S-wave" component is selected for $(V\partial V)^\mu_-$, which vanishes as summarized in the table. Therefore the initial state must be in a ``P-wave" state and $\langle \sigma |v| \rangle_{ann}$ is $v^2$ suppressed.

\section{Doublet/Singlet Scalar as Inelastic Dark Matter} \label{appB}

  \subsection*{Motivation}
  Inelastic dark matter (IDM) remains one of the most promising models to reconcile DAMA and other DM direct detection experiments. But since a relatively large nucleon scattering ($\sigma \geq 2\times10^{-40}\rm cm^2$) is needed to explain DAMA signal, it recently received stringent bound considering neutrinos from DM capture/annhilation in the sun \cite{Menon:2009qj}. The cross-section per nucleon times branching to $WW$ is constrained to be $ \sigma  \lesssim \mathcal{O}(10^{-41}\rm cm^2)$. The constraint is particularly severe with the pure $SU(2)_L$ doublet IDM model considered in \cite{Cui:2009xq}. Combining this with the thermal relic density requirement, we need $\sigma \sim 7\times10^{-39}\rm cm^2$ to fit all data. This shows $O(10^2)$ tension with neutrino bounds. For this reason it is important to ask if IDM may be realized in a different framework.

 In this Appendix we shall propose a different IDM candidate than the scalar mentioned above. We consider a light scalar Dark matter with $M_{DM} < m_W$ that is a doublet/singlet mixture. Our aim is to show that this can be compatible with all the above constraints simultaneously. Futhermore, as described in the the main text, this IDM scenario favors low energy antideuteron production, which is why we bring such detailed attention to it here.

There are several motivations for considering DM lighter than $m_W$. First, according to \cite{Menon:2009qj}, in the case where once two onshell W annihilation channel is closed, the constraint on $\ \sigma$ is relaxed to be $\mathcal{O}(10^{-40} cm^2)$ (unless $\tau$ dominates annihilation). Meanwhile lighter mass implies larger DM flux which allows fitting DAMA data with lower $\sigma$. So the tension between DAMA IDM and solar neutrino bound can be alleviated from two directions. For the case where DM is not much lighter than $m_W$ the dominant channel is $WW^*$ producing  3-body final states (as will be seen later, this is indeed the feature of our model presented here), though we will consider $WW^*$ are almost produced at rest.

Light IDM as a pure $SU(2)$ doublet underproduces dark matter from freezing out of thermal equilibrium. It was already found that doublet scalar (fermion) with mass of $\sim 500 {\rm ~GeV(1 ~TeV)}$ gives right thermal abundance. Roughly, the annihilation cross section scales as\footnote{Not that when $m_{DM}<m_Z/2$, the scaling becomes $m_{DM}^2/m_Z^4$. But for a pure doublet such low mass would hit collider constraint.} $m_{DM}^{-2}$. This monotonic relation implies that it is impossible to get the right relic density for low masses. However an intriguing feature of a light scalar doublet is that
the dominant annihilation channel at freezeout is to light SM fermion pairs via s-channel Z-exchange which has additional ``P-wave" ($v^2$) suppression. This could compensate the increase effect on the early time $\langle \sigma |v|\rangle_{therm}$ arising from lighter masses. From the operator analysis in Appendix A: the operator governing this is $(\phi^\dag\partial^\mu\phi)_-\bar{f}\gamma_\mu(\gamma_5)f$ which ``P-wave" suppressed. But as we will demonstrate by explicit calculation since $v^2\sim1/20$ at freezeout it still cannot provide a sufficient compensation for $m_W<80\rm GeV$. A simple way to fix this is to mix the doublet with a singlet scalar. The necessary mixing angle can be determined by computing relic density of light pure doublet.

Present day $\langle \sigma |v| \rangle_{ann}$ is determined by the off-shell $WW^*$ final state because the phase space suppression is not as large as the velocity suppression for other channels.  When the DM mass is lighter but close to $m_W$, the phase space suppression is not severe so that the 3-body annihilation via $WW^*$ can be dominant channel today with $O(0.01-1)$ suppression compared to $\langle \sigma|v| \rangle_{therm}$ at freezeout. We will explicitly compute $\langle \sigma |v| \rangle_{ann}$ of this 3-body annihilation and find that $m_{DM}=70-80{\rm~ GeV}$ works well for this purpose, and we shall take $m_{DM}\sim 75\rm GeV$ as our benchmark point, which gives $\langle \sigma |v| \rangle_{ann} \sim0.1\langle \sigma |v| \rangle_{therm}$. It is worth mentioning that $v^2$ suppression for all fermion channels also efficiently suppresses the BR to $\tau$ which could potentially impose stronger solar neutrino bound for IDM .

\subsection*{The Model}

    First it is instructive to compute the dominant $\langle \sigma |v| \rangle_{therm}$ at freezeout for a light pure doublet scalar which is annihilating to light SM fermions via Z-exchange
      \beq
           \langle  \sigma |v| \rangle_{therm}=\frac{\lambda_\phi^2}{16\pi}\frac{1}{|( s-m_\phi^2+i m_\phi\Gamma_\phi) |^2}
           \sum_f\lambda_f^2c_f\sqrt{\frac{s-4m_f^2}{s-4m_\phi^2}}\left[\frac{4(s+2m_f^2)(s-4m_\phi^2)}{3s}\right]
       \eeq
where for doublet $\lambda_\phi=\frac{g_2}{2\cos\theta_w}$, $\lambda_f=T^3_f-Q_f\sin^2\theta_w$. The sum is over all light SM fermions. To a good approximation we set all $m_f=0, \Gamma=0$, which leads to
     \beq
          \langle \sigma |v| \rangle_{therm}=\frac{\lambda_\phi^2}{16\pi}\frac{1}{|( s-m_\phi^2+i m_\phi\Gamma_\phi) |^2}
           \sum_f\lambda_f^2c_f\sqrt{\frac{s-4m_f^2}{s-4m_\phi^2}}\left[\frac{4(s+2m_f^2)(s-4m_\phi^2)}{3s}\right]\label{ann1}
     \eeq

We then estimate relic density using\cite{Jungman:1995df}:
     \beqa
     Y^{-1}_\infty&=&0.264g_*^{1/2}m_pm_\phi 3b/(x_f^2)\\\nonumber
     \Omega_\phi h^2&\approx&2.82\times10^8Y_\infty(m_\phi/\rm GeV)
     \eeqa
      where $b$ is the ``P-wave" component from $ \langle \sigma |v| \rangle_{therm}$, which is the coefficient of $v^2$ in eq.(\ref{ann1}) after expanding for small $v^2$. For the mass range we are interested in, it is reasonable to take $h=0.7, x_f=27, g_*=9$. We for DM of masses of $70-80 ~{\rm GeV}$, this pure doublet gives $\Omega h^2 =0.01$. Therefore as discussed earlier, to make critical thermal relic density we need to turn on mixing with singlet to reduce cross section by a factor of $\sim10$. This implies that  converting this model to a thermal relic requires mixing in a singlet candidate is mostly singlet with mixing angle $\sin\theta\sim(0.05-0.1)^{1/4}\sim0.5$. We will now construct such a model.

%\begin{figure}
%\begin{center}
%        \includegraphics[width = 0.4\textwidth]{relic-density}
%\label{relic}
%\end{center}
%\caption{y-axis: relic density $\Omega_\phi$, x-axis: $m_\phi$}
%\end{figure}

     Consider a SM singlet scalar $S$ mixing with an $SU(2)_L$ doublet $D$ with hypercharge $Y=-1$ via SM Higgs. We assume the inelastic splitting between real/imaginary components of the scalar is explained by other physics mechanism (for instance in \cite{Cui:2009xq}) which decouples from our discussion here. We require a $U(1)_{DM}$ under which SM is uncharged, while $S,D$ are charged to forbid large inelastic splitting from term like $S^2h^2$. This combines with proper mass inputs ensure stability of $S$ despite of the vertex $DhS^*$. For simplicity we set input coefficient for possible term $|S|^2h^2$ to be 0 at tree level. Radiatively generated terms of this kind are expected to have negligible effects by combining loop factors and Higgs Yukawa suppressions. The Lagrangian is:
      \beq
       \mathcal{L}=|\partial_\mu S|^2-m_S^2|S|^2+|\mathcal{D}_\mu D|^2-m_D^2|S|^2+(a DhS^*+h.c..)
      \eeq
      where $a$ is a dim-1 parameter (such mixing is similar to what we used in \cite{Cui:2009xq}). The mass matrix to be diagonalized:
     \beq \left(
        \begin{array}{cc}
          S^* & D \\
        \end{array}
      \right)\left(
               \begin{array}{cc}
                 m_S^2 & av \\
                 av & m_D^2 \\
               \end{array}
             \right)
             \left(
               \begin{array}{c}
                 S \\
                 D^* \\
               \end{array}
             \right)
      \eeq
      Mass eigenstates can be written as ($\tilde{S}$ will be our DM candidate, $\theta$: mixing angle)
      \beqa
      \tilde{S}&=&\cos\theta S-\sin\theta D\\
      \tilde{D}&=&\sin\theta S+\cos\theta D
      \eeqa
      Analytic solutions for mass eigenvalues and mixing angle are:
      \beqa
      m_1^2&=&\frac{1}{2}(m_D^2+m_S^2-\sqrt{m_D^4-2m_D^2m_S^2+m_S^4+4a^2v^2})\\
      m_2^2&=&\frac{1}{2}(m_D^2+m_S^2+\sqrt{m_D^4-2m_D^2m_S^2+m_S^4+4a^2v^2})\\
      \tan\theta&=&\frac{1}{2av}(m_D^2-m_S^2+\sqrt{m_D^4-2m_D^2m_S^2+m_S^4+4a^2v^2})
      \eeqa
      We require:\\
      1. $\tilde{S}$ be DM candidate with $m_1<80\rm GeV$\\
      2. We require $\sin\theta\sim0.5$ to get the right relic density\\
      4. We require all input parameters to be Electro-weak scale\\
      A benchmark point that satisfies all the above requirements is:
      $a=15\rm GeV, m_S=85\rm GeV, m_D=100\rm GeV$, this input set gives\\ $m_{DM}\equiv m_1\approx75\rm GeV, m_2\approx108\rm GeV, \sin\theta\approx0.5$\\

      We have checked the compatibility of this dark matter candidate as IDM which explains DAMA data and other DiDt bounds along the lines of \cite{Cui:2009xq}, and find it fits at 99\%CL with $\sigma \approx3\times10^{-40}\rm cm^2$. Such scattering rate is also more compatible with the solar neutrino bound which is $\sim 10^{-40}\rm cm^2$ once highly energetic final state W's are forbidden  and the $\tau$ channel has negligible annihilation fraction. Meanwhile, as mentioned earlier, we find this benchmark point gives $\langle \sigma |v| \rangle_{therm}\sim0.1\rm pb$ today with $WW^*$ final states.

%%%%%%%%%%%%%%%%%%%%%%%%%%%%%%%%%%%%%%%%%%%%%%%%%%%%%%%%%%%%%%%%%

%\newpage

\end{document}